\global\def\draftcontrol{0}

   \def\versionno{Lectures on Statisical Out-of-eq systems -- draft   }
\catcode`\@=11

\expandafter\ifx\csname draftcontrol\endcsname\relax\global\def\draftcontrol{0}
\fi

{\count255=\time\divide\count255 by 60
\xdef\hourmin{\number\count255}
\multiply\count255 by-60\advance\count255 by\time
\xdef\hourmin{\hourmin:\ifnum\count255<10 0\fi\the\count255}}
\def\draftdate{\number\month/\number\day/\number\year\ \ \ \hourmin }

\newcommand\makepapertitle{\par
  \begingroup
    \renewcommand\thefootnote{\@fnsymbol\c@footnote}%
    \def\@makefnmark{\rlap{\@textsuperscript{\normalfont\@thefnmark}}}%
    \long\def\@makefntext##1{\parindent 1em\noindent
            \hb@xt@1.8em{%
                \hss\@textsuperscript{\normalfont\@thefnmark}}##1}%
     \newpage
     \global\@topnum\z@   % Prevents figures from going at top of page.
     \@makepapertitle
     \thispagestyle{empty}\@thanks
  \endgroup
  \setcounter{footnote}{0}%
  \global\let\thanks\relax
  \global\let\makepapertitle\relax
  \global\let\@makepapertitle\relax
  \global\let\@thanks\@empty
  \global\let\@author\@empty
  \global\let\@date\@empty
  \global\let\@title\@empty
  \global\let\title\relax
  \global\let\author\relax
  \global\let\date\relax
  \global\let\and\relax
  \def\version{\let\version\@version\@gobble}
}
\def\@makepapertitle{%
  \newpage
   \ifnum\draftcontrol=1 {}
   \version\versionno
   \vskip 3em%
   \else
   \hfill\hbox to 3cm {\parbox{4cm}{\@pubnum}\hss}%
   \vskip 3em%
   \fi
   \begin{center}%
   \let \footnote \thanks
     {\LARGE {\@title}}%
     \vskip 1.5em%
     {\normalsize%\large
       \lineskip .5em%
       \begin{tabular}[t]{l}%
         \@author
       \end{tabular}\par}%
     \vskip 1.5em%
     {\@bstract}%
     \end{center}%
     \vskip 1.5em
     \@date%
   \par
}

\gdef\@pubnum{}
\def\pubnum#1{%
  \gdef\@pubnum{#1}}

\gdef\@bstract{}
\def\Abstract#1{%
  \gdef\@bstract{%
   \parbox{\textwidth-0pc}{%
   \centerline{\bf Abstract}\penalty1000%
\kern.2cm%
\noindent%\abstractfont \baselineskip=12pt
\renewcommand\baselinestretch{1.0}%
{#1}}}
}

\def\ps@paper{\let\@mkboth\@gobbletwo%
     \ifnum\draftcontrol=1
    \def\@oddfoot{\hbox to \textwidth{\tiny \versionno \hfil\tiny\draftdate}%
    \hskip -\textwidth \hbox to \textwidth{\hfil\rm\thepage\hfil}}%
     \else\def\@oddfoot{\hbox to \textwidth{\hfil\rm\thepage\hfil}}
     \fi
     \let\@evenfoot\@oddfoot
}
\newenvironment{acknowledgements}{%
\vskip 3.25ex
\noindent {\bf Acknowledgements}
}

\def\@version#1{\ifnum\draftcontrol=1
\typeout{}\typeout{#1}\typeout{}
\vskip3mm\centerline{\hbox{\fbox{\normalsize{\tt DRAFT -- #1 -- }
                   {\draftdate}}}}\vskip3mm
\fi}
\let\version\@version
\long\def\eqlabel#1{\ifnum\draftcontrol=1
                    \tag@false  % there are some problems with multline without this
                    \tag*{(\theequation) \hbox to -0.2cm{\hspace{0cm}\small{#1}\hss}}
                    \refstepcounter{equation}
                    \edef\@currentlabel{\theequation}
                    \ltx@label{#1}          % use old LaTeX \label instead of new definition
                    \else
                    \label{#1}
                    \fi
                    }
\let\st@bibitem\@bibitem
\let\st@lbibitem\@lbibitem
\ifnum\draftcontrol=1
  \def\@bibitem#1{%
    \st@bibitem{#1}\a@@label{#1}\ignorespaces}
  \def\@lbibitem[#1]#2{%
    \st@lbibitem[#1]{#2}\a@@label{#2}\ignorespaces}
  \def\a@@label#1{%
    \gdef\a@lab{\smash{\normalfont\small#1}}
    \ifvmode
      \if@inlabel
        \global\setbox\@labels\hbox{%
          \llap{\a@lab\let\a@lab\relax
                \kern\@totalleftmargin\kern\marginparsep}%
          \box\@labels}%
      \fi
    \fi}
\fi
\documentclass[10pt]{article}
\usepackage{amsmath,amssymb,array,calc,epsfig,bm}
\usepackage{graphicx}        % standard LaTeX graphics tool
\usepackage{multicol}        % used for the two-column index

\usepackage{amsmath,amssymb,bm,epsfig}

\usepackage{cite}
\usepackage[
      colorlinks=true,
      linkcolor=red,  
      urlcolor=blue,    
      filecolor=red,     
      linkcolor=blue,
      citecolor = red,
      pdfstartview=FitV,
      pdftitle={},
        pdfauthor={Paolo Benincasa},
        pdfsubject={},
        pdfkeywords={},
        pdfpagemode=None,
        bookmarksopen=true    
      ]{hyperref}

\usepackage{centernot}      
\usepackage{mathpazo}
\usepackage{stmaryrd}
\usepackage{pifont}
\usepackage{colortbl}
\usepackage{multirow,bigdelim}
\usepackage{arydshln}
\usepackage{tikz, pict2e}
\usepackage{geometry}
\usepackage{lmodern}
\usepackage{wrapfig}
\usepackage{float}
\usepackage{enumitem}
\usepackage[framemethod=default]{mdframed}
\usepackage{array}
\usepackage{mathrsfs}

\usepackage{placeins}
\usepackage{circuitikz}

\usepackage[T1]{fontenc}
\usepackage[utf8]{inputenc}
\usepackage[english]{babel}

\PassOptionsToPackage{table}{xcolor}

\usepackage{xspace}
\usepackage{ifthen}
\usepackage{epstopdf}
\usepackage{amsthm}

\usepackage{relsize}
\usepackage{tabularx}
\usepackage{booktabs}
\usepackage{soul}
\usepackage{pdfpages}
\usepackage[most]{tcolorbox}
\usepackage{capt-of}
\usepackage{xfrac}
\usepackage{empheq}
\usepackage{slashed}
\usepackage{cancel}
\usepackage{upgreek}
\usepackage[skip=1em]{parskip}

\usepackage{latexsym}
\usepackage{ifoddpage}
\usepackage{subcaption}

\usepackage{cancel}
\usepackage{comment}
\usepackage{shuffle}

\usepackage{savesym}
\savesymbol{Ref}
\savesymbol{erf}
\usepackage{physics}
\restoresymbol{phys}{Ref}
\restoresymbol{phys}{erf}

\usepackage[compat=1.1.0]{tikz-feynman}
\tikzfeynmanset{compat=1.1.0}

\definecolor{myblue}{RGB}{166, 193, 222}
\definecolor{myblue2}{RGB}{68, 114, 196}
\definecolor{mygreen}{RGB}{175, 218, 162}
\definecolor{mygreen2}{RGB}{133, 210, 110}
\definecolor{myorange2}{RGB}{255, 192, 0}

\newtcolorbox{shade}[1][]{
  top=-7pt,bottom=6pt,
  sharp corners=all,
  colback=white!95!black,
  colframe=white,
}

\newcolumntype{P}[1]{>{\centering\arraybackslash}p{#1}}
\newcolumntype{L}{>{$}l<{$}}

\usetikzlibrary{calc,backgrounds}
\usetikzlibrary{arrows,shapes,positioning,shadows,trees}
\usetikzlibrary{arrows.meta, shapes.geometric}
\usetikzlibrary{mindmap}
\usetikzlibrary{decorations.pathreplacing}
\usetikzlibrary{decorations.pathmorphing}
\usetikzlibrary{decorations.markings}
\usetikzlibrary{matrix}
\usetikzlibrary{patterns.meta}

\tikzstyle arrowstyle=[scale=1]
\tikzstyle directed=[postaction={decorate,decoration={markings,
    mark=at position .5 with {\arrow[arrowstyle]{stealth}}}}]
\tikzstyle reverse directed=[postaction={decorate,decoration={markings,
    mark=at position .5 with {\arrowreversed[arrowstyle]{stealth};}}}]

\ifnum\draftcontrol=1
\fi
\renewcommand\baselinestretch{1.25}
\renewcommand\section{\@startsection {section}{1}{\z@}%
                                   {-3.5ex \@plus -1ex \@minus -.2ex}%
                                   {2.3ex \@plus.2ex}%
                                   {\normalfont\large\bfseries}}
\renewcommand\subsection{\@startsection{subsection}{2}{\z@}%
                                   {-3.25ex\@plus -1ex \@minus -.2ex}%
                                   {1.5ex \@plus .2ex}%
                                   {\normalfont\normalsize\bfseries}}
\renewcommand\subsubsection{\@startsection{subsubsection}{3}{\z@}%
                                   {-3.25ex\@plus -1ex \@minus -.2ex}%
                                   {1.5ex \@plus .2ex}%
                                   {\normalfont\normalsize\it}}
\renewcommand\paragraph{\@startsection{paragraph}{4}{\z@}%
                                   {-3.25ex\@plus -1ex \@minus -.2ex}%
                                   {1.5ex \@plus .2ex}%
                                   {\normalfont\normalsize\bf}}

\numberwithin{equation}{section}

\def\revise#1       {\raisebox{-0em}{\rule{3pt}{1em}}%
                     \marginpar{\raisebox{.5em}{\vrule width3pt\
                     \vrule width0pt height 0pt depth0.5em
                     \hbox to 0cm{\hspace{0cm}{%
                     \parbox[t]{4em}{\raggedright\footnotesize{#1}}}\hss}}}}

\def\sqr#1#2{{\vcenter{\vbox{\hrule height.#2pt
 \hbox{\vrule width.#2pt height#1pt \kern#1pt
 \vrule width.#2pt}\hrule height.#2pt}}}}

\catcode`\@=12

\newcommand{\ee}{e}

\newcommand{\beq}{\begin{equation}}
\newcommand{\eeq}{\end{equation}}
\newcommand{\bea}{\begin{equation}\begin{aligned}}
\newcommand{\eea}{\end{aligned}\end{equation}}

\newlength{\wsingfig}
\newlength{\wdblefig}
\newlength{\wquadfig}
\newlength{\wtriplefig}
\newcommand   \lptl{\raise .8ex\hbox{$^\leftarrow$} \hspace{-9pt} \partial}
\newcommand   \lrptl{\raise .8ex\hbox{$^\leftrightarrow$} \hspace{-9pt} \partial}

\newcommand   \be {\mathbf{e}}

\def\be#1\ee{\begin{align}#1\end{align}}

\newcounter{ssooeexercise}[section]
\renewcommand{\thessooeexercise}{\thesection.\arabic{ssooeexercise}}

\newtcolorbox{exbox}{
  colback=gray!10,
  colframe=gray!20,
  boxrule=1pt,
  arc=5pt,
  breakable,
  left=1em,
  right=1em,
  top=0.8em,
  bottom=0.8em
}

\newcommand{\newexercise}[3][]{%
  \refstepcounter{ssooeexercise}%
  \expandafter\gdef\csname ex@content@#2\endcsname{#3}%
  \expandafter\xdef\csname ex@number@#2\endcsname{\thessooeexercise}%
  \begin{exbox}%
    \textbf{Exercise~\thessooeexercise:} #3
    \ifx&#1&\else\label{#1}\fi
  \end{exbox}%
}

\newcommand{\printexercise}[1]{%
  \begin{exbox}%
    \textbf{Exercise~\csname ex@number@#1\endcsname:}
    \csname ex@content@#1\endcsname
  \end{exbox}%
}

\begin{document}

%%%%%%%%%
%%%%%% text starts here
%%%%%%%%%

\title{\underline{\Large\bf The Disordered Universe 2025:} \\
        {\Large\bf Statistical Systems Out of Equilibrium}}

\pubnum{%
}
  
\author{%
	{\scshape Lecturer:} {\bf Tarek Anous}${}^{\circledast}$,%
    \\
    {\scshape Notes written by:} {\bf Weam Abou Hamdan}${}^{\star}$, {\bf Mart{\'i} Berenguer Mim{\'o}}${}^{\ddagger\,\circledcirc}$, {\bf Rafael Carrasco Carmona}${}^{\ast}$, \\
      \hspace{3.375cm} {\bf Silvia Pla}${}^{\odot}$, {\bf Omkar Shetye}${}^{\oplus}$
    \\
    {\scshape Edited by:} {\bf Paolo Benincasa}${}^{\dagger\,\ddagger}$, {\bf Guilherme Leite Pimentel}${}^{\diamond}$
    \\ [0.4cm]
    \ttfamily ${}^{\circledast}$ School of Mathematical Sciences, Queen Mary University of London, \\
    \ttfamily $\phantom{{}^{\dagger}}$ Mile End Road, London, E1 4NS, UK\\
    \ttfamily ${}^{\star}$ Department of Mathematics, \\
    \ttfamily $\phantom{{}^{\dagger}}$ King’s College London, Strand, London WC2R 2LS, UK \\
    \ttfamily ${}^{\ddagger}$Departamento de Física de Partículas, Universidade de Santiago de Compostela and \\ 
    \ttfamily $\phantom{{}^{\ddagger}}$ Instituto Galego de Física de Altas Enerxías (IGFAE), E-15782 Galicia-Spain\\
    \ttfamily ${}^{\circledcirc}$ Asia Pacific Center for Theoretical Physics, \\ 
    \ttfamily $\phantom{{}^{\ddagger}}$ 7673 Pohang, Gyeongsangbuk-do, South Korea\\
    \ttfamily ${}^{\ast}$ Instituto de F{\'i}sica Te{\'o}rica UAM/CSIC,\\
    \ttfamily $\phantom{{}^{\dagger}}$ Calle Nicol{\'a}s Cabrera 13-15, Cantoblanco, 28049 Madrid, Spain  \\
    \ttfamily ${}^{\odot}$ Physik-Department, Technische Universit{\"a}t M{\"u}nchen, \\
    \ttfamily $\phantom{{}^{\dagger}}$ James-Franck-Str., 85748 Garching, Germany\\ 
    \ttfamily ${}^{\oplus}$ Instytut Fizyki Teoretycznej, \\
    \ttfamily $\phantom{{}^{\dagger}}$ Uniwersytet Jagielloński, Lojasiewicza 11, 30-348 Kraków, Poland \\
    \ttfamily ${}^{\dagger}$ Asociaci{\'o}n Cultural La Ricotta, Espazo La Ricotta,\\
    \ttfamily $\phantom{{}^{\dagger}}$ As Barreiras, Castro Caldelas, E-32769 Galicia-Spain\\
    \ttfamily ${}^{\diamond}$ Scuola Normale Superiore, \\
    \ttfamily $\phantom{{}^{\dagger}}$ Piazza dei Cavalieri 7, 56126, Pisa, Italy\\[0.4cm]
    \small \ttfamily t.anous@qmul.ac.uk, {\textsf{{\nolinkurl{weam.abou\_hamdan@kcl.ac.uk}}} }, marti.berenguer.mimo@usc.es, \\
      \small \ttfamily rafael.carrasco@ift.csic.es, silvia.pla-garcia@tum.de, omkar.sanjay.shetye@uj.edu.pl,\\    
      \small \ttfamily pablowellinhouse@sindominio.net, guilherme.leitepimentel@sns.it \\ [0.75cm]
}

\Abstract{These lecture notes provide an introduction to quantum spin glasses, with a particular emphasis on their application to open problems in cosmology and holography. Focusing on the spherical $p$-spin model, we develop a complete and self-contained analysis of its thermodynamic and dynamical properties. We introduce the replica method, discuss both replica-symmetric and one-step replica symmetry-broken solutions, and characterise the resulting glassy phase. The framework is then extended to the quantum model, where we study the emergence of slow dynamics and the role of quantum correlations through Euclidean and real-time correlation functions. Finally, we explore how concepts such as replica symmetry breaking and disorder naturally arise in cosmological and holographic systems related to black hole fragmentation, highlighting spin glass models as useful theoretical laboratories for these phenomena. These notes are based on a series of lectures held during the \href{https://indico.sns.it/event/91/}{{\it La Ricotta} Summer School: {\it The Disordered Universe 2025}} from 21~July to 1~August~2025.}

\makepapertitle

%\body

\version\versionno

\newpage

\section*{Preface}
This article is a chapter from the book {\it The Disordered Universe: Open Systems, Scattering Amplitudes, de Sitter Space, and Disordered Systems in Cosmology}~\cite{Benincasa:2027tdu}, which collects the lectures delivered at the \href{https://indico.sns.it/event/91/}{\it La Ricotta Summer School: The Disordered Universe 2025}, held from 21~July to 1~August~2025 at the \href{https://blogs.sindominio.net/laricotta/#}{\it Espazo La Ricotta} in As Barreiras (Castro Caldelas, Spain). These lecture notes are intended as a pedagogical introduction to topics that are becoming increasingly relevant in cosmology. 

\newpage

\tableofcontents

\section{Introduction and motivation}\label{sec:intro-ssooe}

One of the most familiar phase transitions in everyday life is the transition from a liquid to a glass. Unlike crystallization, which produces a solid with long-range order, the glass transition leads to a rigid configuration that lacks any obvious spatial pattern. Microscopically, the constituents of a glass are frozen in disordered configurations: the system loses its ability to flow, and it evades the crystalline structure.

This distinction is crucial. The glass transition underlies technologies as diverse as optical lenses, cryogenic preservation of biological material, and even the texture of everyday food such as ice cream. Despite its practical importance in all aspects of life, the glass transition remains one of the most challenging open problems in condensed matter physics.

From a theoretical perspective, phase transitions are typically characterized by sharp changes in behavior, often associated with the emergence of an order parameter or the divergence of a characteristic length or timescale. The difficulty with the glass transition is that no such order parameter is known. Realistic models of glass-forming liquids are complicated, strongly-interacting many-body systems, and this complexity obscures a clean theoretical characterization of the transition.

An alternative approach is to study {\it spin glass models}, which provide simplified yet highly non-trivial examples of disordered systems with slow dynamics. Although spin glasses are not literal models of structural glasses, they capture several essential features, such as frustration and the emergence of phases characterized by frozen disorder rather than conventional symmetry breaking.

The purpose of these lectures is to present a detailed and self-contained study of one such model: the spherical $p$-spin glass \cite{Crisanti:1992sxo,Cugliandolo_2000,Castellani_2005,Cugliandolo_2001}. We will analyze this model from first principles, introducing the replica method, mean-field theory, and replica symmetry breaking, and we will follow the system from its high-temperature paramagnetic phase into a genuinely glassy regime.

Beyond their original condensed matter context, spin glasses have found striking applications in other areas of theoretical physics. In recent years, techniques developed in the study of disordered systems have played an important role in quantum gravity, holography, and cosmology. The appearance of the SYK model \cite{Sachdev:1992fk,Kitaev:2015,Maldacena:2016hyu,Kitaev:2017awl} provided a concrete bridge between disordered quantum systems and low-dimensional gravity. In the low energy-limit, its dynamics is governed by a soft mode described by the Schwarzian action, which at the same time describes the only degree of freedom of JT gravity \cite{Jackiw_1984,Teitelboim_1983,Maldacena_2016}. The spherical $p$-spin model, in particular, shares deep structural similarities with the SYK model and many variants of it, which arise in the study of black hole dynamics, AdS fragmentation \cite{Maldacena:1998uz}, and the quantum mechanics of expanding universes \cite{Anninos:2011kh}.

These connections provide an additional motivation for the present course. By working through a concrete and solvable model of a spin glass in detail, we aim not only to illustrate the physics of glassy systems, but also lay the groundwork for understanding how similar ideas emerge in cosmology and holography.

The structure of these notes is as follows. In Section \ref{sec:meanfield}, we introduce the mean-field approximation in a simplified model of ferromagnetism. Section \ref{sec:sphericalpspin} is devoted to the study of the spherical $p$-spin model using the replica method and replica symmetry and its breaking in the static approximation. In Section \ref{sec:dynamics}, we study the quantum dynamics of the model and its real-time behavior. Finally, in Section \ref{sec:relationcosmo} we discuss how the concepts and techniques developed throughout these notes connect to problems in cosmology and holography. Appendix \ref{app:solutions} contains the solutions to the exercises proposed throughout the main text.

\section{Mean-field spin glass models}\label{sec:meanfield}

\subsection{How do we characterize a phase transition? A lattice example}

Before turning to spin glasses, it is useful to recall how phase transitions are characterized in simpler systems. Broadly speaking, there are two complementary viewpoints: {\it statics} and {\it dynamics}. In this section we focus on the static case, using a mean-field ferromagnetic model as an illustrative example.

\begin{figure}[t]
	\centering
	\includegraphics{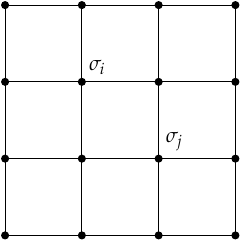}
\caption{Classical Ising lattice with spins $\sigma_i$, $i=1,\dots,N$.}
	\label{fig:Isinglattice}
\end{figure}

\paragraph{\bf A spherical lattice model.}
Consider a classical model of $N$ `spins' living on a lattice as in Fig. \ref{fig:Isinglattice}, with Hamiltonian
\begin{equation}
    H = -\frac{J}{2N} \sum_{\left\{ i,j \right\}}\, \sigma_i \sigma_j \ ,
\end{equation}
where $N$ denotes the number of sites. The sum $\left\{i,j\right\}$ may represent nearest neighbors, but we will eventually relax this requirement and treat the interaction as all-to-all. This is the \emph{mean-field} approximation; see, for example, Refs.~\cite{Binney:1992vn,Cardy_1996}. The `spins' are not quite spins, but will be treated as continuous variables,
\begin{equation}
    -\infty <\sigma_i<\infty\, , \quad \forall i=1,\dots,N\, .
\end{equation}
This Hamiltonian is unfortunately unbounded from below, so to rectify this we must supplement it with a {\it spherical constraint}~\cite{Berlin:1952zz}
\begin{equation}\label{eq: spherical constraint}
    \sum_i \sigma_{i}^2=N\, .
\end{equation}
This model captures the essence of ferromagnetic ordering while remaining analytically
tractable. Since we are working in \emph{mean-field}, where interactions are
assumed to be of infinite range, spatial fluctuations are suppressed. Nevertheless this model is good for developing a clear picture for characterizing the ferromagnetic phase transition.

\paragraph{\bf Partition function and constraints.}
We are interested in the partition function at inverse temperature $\beta$ and in the presence
of an external magnetic field $h$,
\begin{equation}\label{eq:partitionfunctionlattice}
    Z_N[\beta,h]=\frac{1}{2\pi}\int_{-\infty}^{\infty}\text{d}z \int_{-\infty}^{\infty}\text{d}\sigma_1\cdots \int_{-\infty}^{\infty}\text{d}\sigma_N\exp\left(\frac{\beta J}{2N} \sum_{i\neq j}\sigma_i\sigma_j+(a+iz)(N-\sum_i\sigma_i^2)+h \sum_i \sigma_i \right)\, .
\end{equation}
To handle the constraint \eqref{eq: spherical constraint}, we recall the integral representation of the $\delta$-function,
\begin{equation} \label{eq:lecture1:identities01}
\delta(x)=\frac{1}{2\pi}\int_{-\infty}^{\infty}\text{d}z e^{ixz}\, ,
\end{equation}
and we also note that inserting
\begin{equation}
    \qquad a(N-\sum_{i=1}^{N}\sigma_i^2)=0
\end{equation}
inside the exponential is harmless and will prove convenient later.

We can simplify the expression above by noticing the following property:
\begin{equation}
    \sum_{i\neq j}\sigma_i \sigma_j=\big(\sum_i \sigma_i\big)^2-\sum_i \left( \sigma_i^2\right)\, ,
\end{equation}
which allows to perform the integral over the spins,
\begin{equation}
    \exp\left(\frac{\beta J}{2N}\big(\sum_i \sigma_i\big)^2\right)=\sqrt{\frac{\beta J N}{2\pi}}\int_{-\infty}^{\infty}\text{d}m \exp\left(\frac{N\beta J}{2}m^2+\beta J m \sum_i \sigma_i\right)\, ,
\end{equation}
where $m$ is introduced as an auxiliary variable. The integrand has a peak at $m\approx N^{-1}\sum_i\sigma_i$, so $\langle m \rangle$ will be identified with the average magnetization.

Applying these results to the partition function \eqref{eq:partitionfunctionlattice}, we obtain
\begin{equation}
\begin{aligned}  
    Z_N[\beta,h]&=\frac{1}{2\pi} \sqrt{\frac{\beta J N}{2\pi}}\int_{-\infty}^{\infty}\text{d}z \int_{-\infty}^{\infty}\text{d}m \int_{-\infty}^{\infty}\text{d}\sigma_1\cdots \int_{-\infty}^{\infty}\text{d}\sigma_N \exp\Bigg( -\frac{N \beta J}{2}m^2 + (\beta J m + h)\sum_i\sigma_i\\  &\qquad\qquad\qquad\qquad\qquad\qquad\qquad\qquad\qquad - \frac{\beta J}{2N}\sum_i \sigma_i^2+(a+iz)N-(a+iz)\sum_i \sigma_i^2\Bigg)\, .
    \end{aligned}
\end{equation}

\begin{figure}[t]
    \centering   \includegraphics[width=0.65\linewidth]{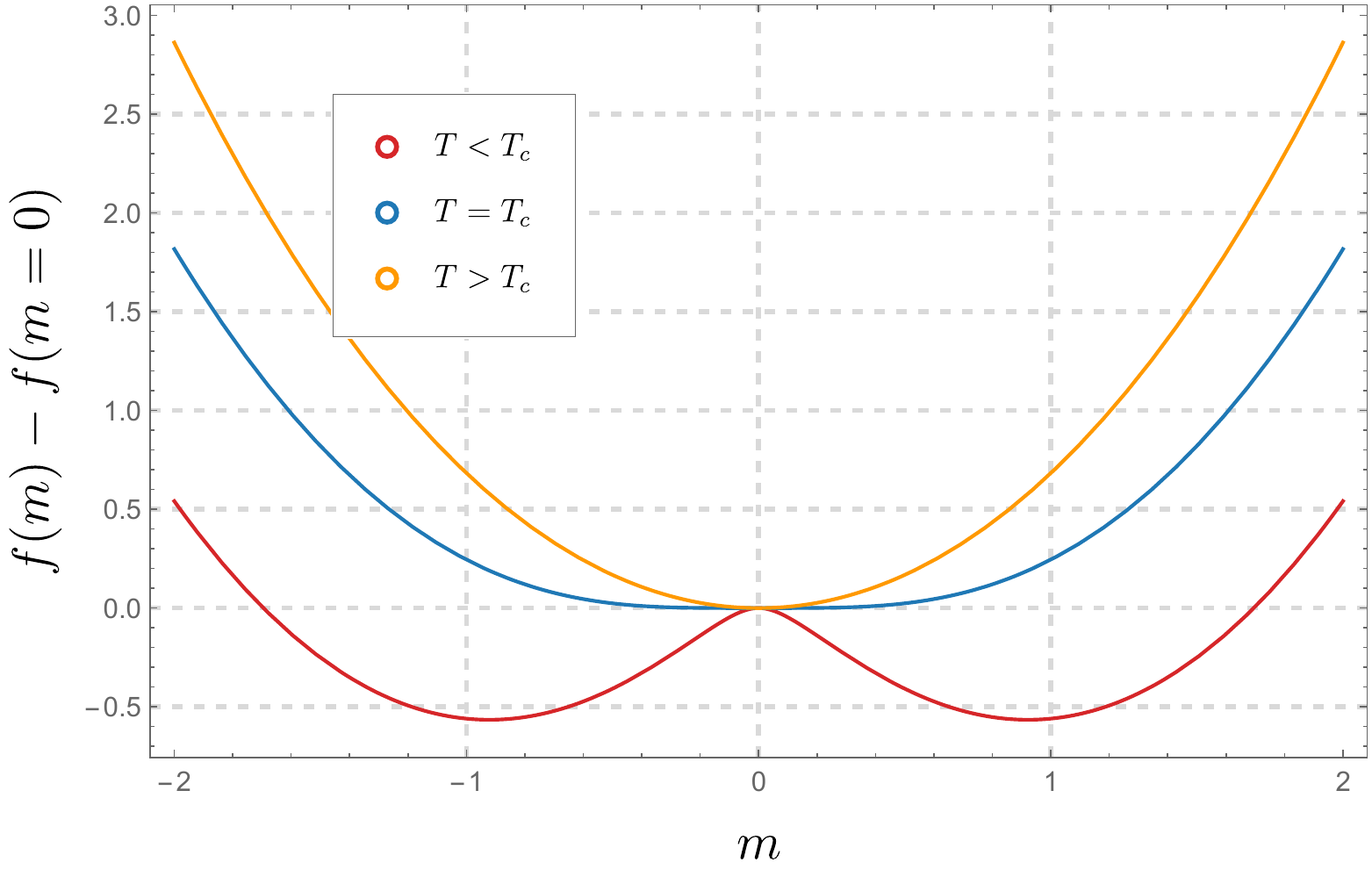}
    \caption{Free energy as a function of $m$ for three different values of the temperature.}
    \label{fig:free_energy_m}
\end{figure}

\paragraph{\bf Large-$N$ limit and free energy.}
At large-$N$, the term $\frac{\beta J}{2N}\sum_i \sigma_i^2$ in the exponential is subleading. The $N$ integrals over the spins are Gaussian and can be performed explicitly, yielding
\begin{equation}
\begin{aligned} \label{eq:ZN-zm01}
    Z_N[\beta,h]=\frac{1}{2\pi}\sqrt{\frac{N \beta J}{2\pi}} \int_{-\infty}^{\infty}\text{d}z \int_{-\infty}^{\infty}\text{d}m\exp
\Bigg(\frac{N}{2}\Bigg(&-\beta J m^2 + \frac{1}{2}\frac{(h+\beta Jm)^2}{a+iz +\frac{\beta J}{N}}\\
   & +2(a+iz)-\log \frac{a+iz+\frac{\beta J}{N}}{\pi}\Bigg)\Bigg)\, .
    \end{aligned}
\end{equation}
In the large-$N$ limit, we can use the saddle-point approximation and evaluate the partition function on-shell (see, for example Ref. \cite{Bender:1999box}). The saddle-point reads
\begin{equation}
    \delta_z:\qquad  z_*=i\left[\left(a+\frac{\beta J}{N}\right)-\frac{1}{4}\left(1+2\beta J \sqrt{\left(m+\frac{h}{\beta J}\right)^2+\frac{1}{4\beta^2J^2}}\right) \right]\, ,
\end{equation}
which we can plug back into $Z_N$ evaluated at $h=0$, and find 
\begin{equation}\label{eq:free_energy_part_function}
    Z_N\approx \frac{1}{2\pi }\sqrt{\frac{N\beta J}{2\pi}}\int_{-\infty}^{\infty} \text{d}m\exp{-N\beta \,f}\, ,
\end{equation}
where $f$ is the free energy per site,
\begin{equation} \label{eq:freee}
    f(m,T,J)=\frac{1}{2}\left\{m^2 J-T-\frac{4J^2m^2}{T+\sqrt{T^2+4J^2m^2}}+T \log \left(\frac{1+\sqrt{1+\frac{4J^2m^2}{T^2}}}{4\pi}\right)\right\}\, .
\end{equation}
Fig. \ref{fig:free_energy_m} shows the free energy as a function of $m$ for several values of the
temperature. As the temperature is lowered, the free energy changes from having a single minimum to showing two symmetric minima, corresponding to spontaneous magnetization.

\newexercise[ex:integrateoutz]{ex:integrateoutz}{Integrate out $z$ at large $N$ and obtain the $f[m,T,h]$ given in \eqref{eq:freee}. Study its behavior.}

\begin{figure}[t]
    \centering   \includegraphics[width=0.65\linewidth]{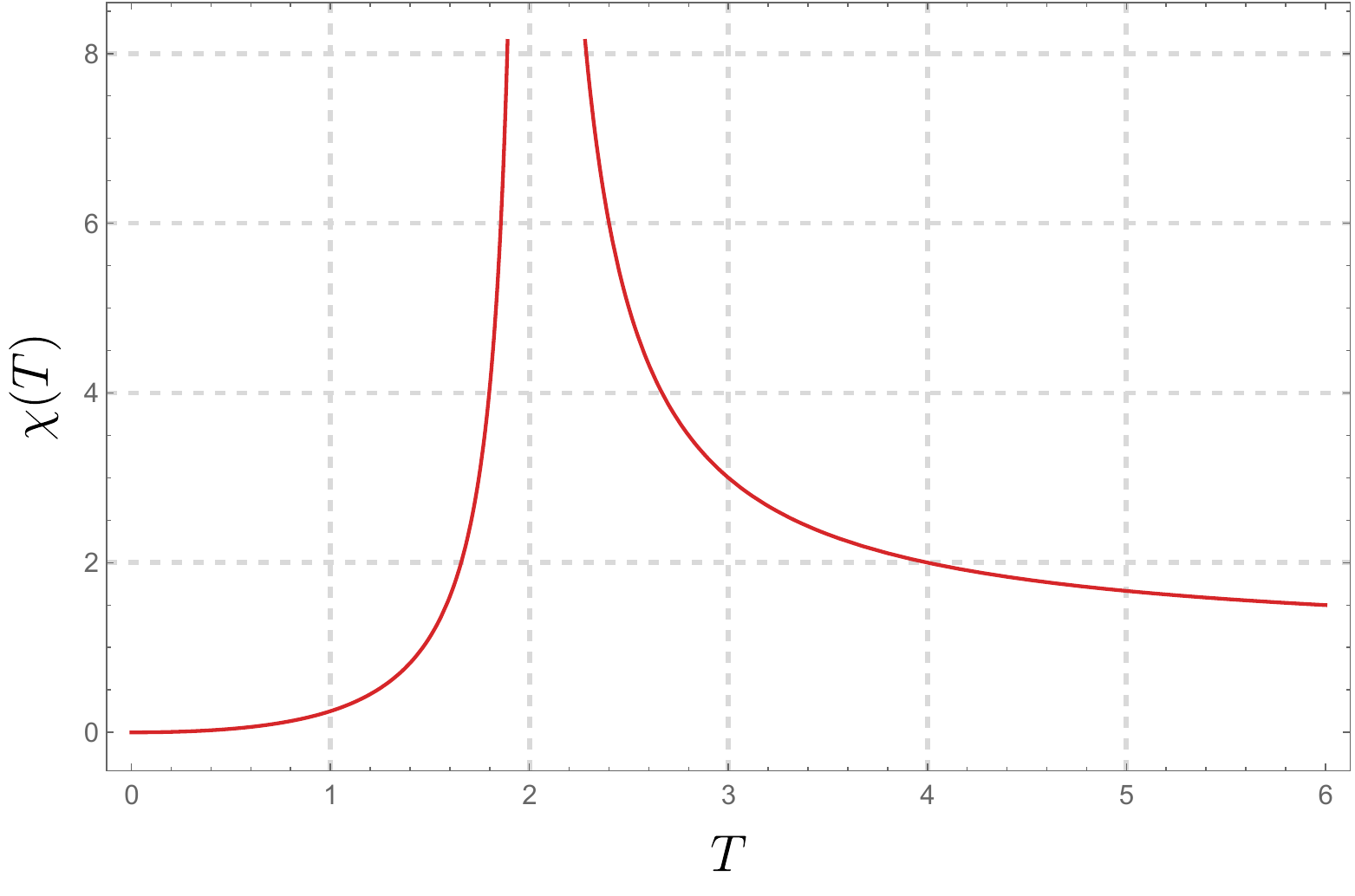}
    \caption{Susceptibility $\chi$ as a function of the temperature $T$.}
    \label{fig:sus}
\end{figure}

\paragraph{\bf Susceptibility and critical behaviour.}
The phase transition can be characterized more sharply by studying the magnetic susceptibility, $m(h,T)$,
which measures the response of the magnetization to an external field. Defining
\begin{equation}
    \langle m \rangle =\lim_{h\to 0^{+}}\langle m(h,T) \rangle\, ,
\end{equation}
the susceptibility is
\begin{equation}
\label{eq:susceptibility}
\chi=\left.\frac{\partial m\left(h, T\right)}{\partial h}\right|_{h=0}=\left\{\begin{array}{l}
\alpha_1 \frac{T^2}{\left(1-\frac{T}{T_c}\right)}\qquad  T <T_c \\
\\
\alpha_2 \frac{T}{\left(\frac{T}{T_c}-1\right)} \qquad T>T_c
\end{array}\right.
\end{equation}
At the critical temperature $T_c$, the susceptibility diverges, indicating that the system is
extremely sensitive to external perturbations. We show this behavior in Fig. \ref{fig:sus}.

\newexercise[ex:susceptibility]{ex:susceptibility}{Prove the expression \eqref{eq:susceptibility} for the susceptibility and determine the coefficients $\alpha_1$ and $\alpha_2$.}

\paragraph{\bf Beyond mean field.}
More generally, the magnetization can be written as
\begin{equation}
    \langle m\rangle =\frac{1}{Z} \sum_{\text{configuration}}\sigma_0 e^{-\beta H(\sigma)+h \sum_i \sigma_i}\, ,
\end{equation}
and the susceptibility is given by
\begin{equation}
   \frac{\partial \langle m\rangle}{\partial h}\Bigg|_{h=0} =\sum_i \langle \sigma_i \sigma_0\rangle - \langle\sigma_0^2 \rangle %\longrightarrow_\text{continuum}
   \underset{\text{continuum}}{\longrightarrow}\quad \int \text{d}^dx~G(\vec x) \, 
\end{equation}
where $G(\vec x)$ is the connected correlation function. Under mild assumptions (translation invariance, homogeneity, isotropy and locality),
\begin{equation}
    G(\vec x)\sim \frac{e^{-|\vec x|/\xi}}{|\vec x|^{1+\eta}},
\end{equation}
so that 
\begin{equation}
    \frac{\partial m}{\partial h}\Bigg|_{h=0}\propto \xi^{d-1-n}\, .
\end{equation}
In other words, we see that the divergence of the susceptibility is associated with a divergence of the correlation length $\xi$ \cite{Binney:1992vn,Cardy_1996}. From this perspective, mean-field theory corresponds to the limit $d=\infty$.

\newexercise[ex:FSandE]{ex:FSandE}{Compute the free energy $F$, entropy $S$, energy $E$ and all thermo of this model. More concretely, show that
$$
\frac{F}{N}= \begin{cases}a+b\, T \log \frac{T}{c} & T<T_c \\ d \, T & T>T_c\end{cases}
$$
and 
$$
C_T=\frac{\partial\langle E\rangle}{\partial T} \begin{cases}\# & T<T_c \\ 0 & T>T_c\end{cases}
$$
where $C_T$ is the specific heat. Determine $a$, $b$, $c$ and $d$. Try via  saddle point vs exact integration, what do you notice?\\
\textbf{Hint:} It is useful to remind the structure of the partition function $Z=\sum_{\text{conf.}}e^{-\beta H}$. The specific heat is defined as  $$C_T\equiv \frac{\partial E}{\partial T}\, ,$$
where
$$\langle E \rangle=\frac{1}{Z} \sum_{\text{configs.}}H e^{-\beta H}=-\partial_\beta \log Z=\partial_\beta(\beta F)=-T^2\partial_T\left(\frac{F}{T}\right)=F-T\partial_TF\, .$$
The free energy reads $F=\langle E\rangle +T\partial_T F$. But recall $F=\langle E\rangle-T\langle S \rangle$ so that $\langle S\rangle =-\partial_T F$.}

\paragraph{\bf Towards glasses.}

\begin{figure}[t]
    \centering
    \includegraphics[width=0.65\linewidth]{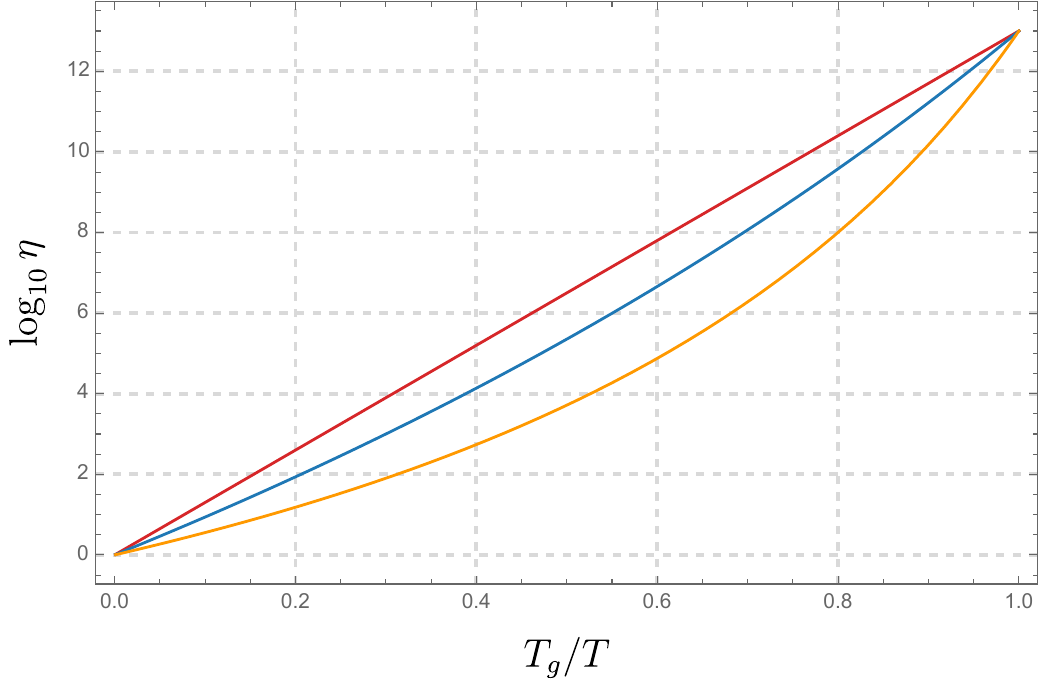}
    \caption{Shear viscosity $\eta$ as a function of the temperature. Super-exponential increase in viscosity as temperature is lowered.}
    \label{fig:shearV}
\end{figure}

While the ferromagnetic transition is characterized by a diverging order parameter and susceptibility, glass formers behave differently. Their defining feature is extremely slow dynamics: quantities such as the shear viscosity grow rapidly as the temperature is lowered,
without exhibiting a sharp thermodynamic singularity (see Fig. \ref{fig:shearV}). This motivates the search for models where freezing occurs without conventional long-range order, a theme to which we now turn (see Ref. for a pedagogical review  \cite{Castellani_2005}).

\section{The spherical \texorpdfstring{$\boldsymbol{p}$}{p}-spin model}\label{sec:sphericalpspin}

In a real glass, the positions of the constituent particles are effectively frozen, much as in a
crystal. However, unlike a crystal, there is no underlying periodic order: the frozen configuration
is amorphous. Our goal is to construct a simple statistical model that captures this essential
feature. 

Motivated by the ferromagnetic example studied previously, we again consider continuous spin variables $\sigma_i$. This time, however,  the local magnetizations $\sigma_{i}$ will be fixed, but not in any specific direction. In particular, even at zero temperature the average magnetization should vanish,
\begin{equation}
    m=\frac{1}{N}\sum_i \langle \sigma_i\rangle =0\, ,
\end{equation}
while the system itself remains frozen. In other words, the system should be frozen but not ordered.

\begin{figure}[t]
	\centering
	\includegraphics{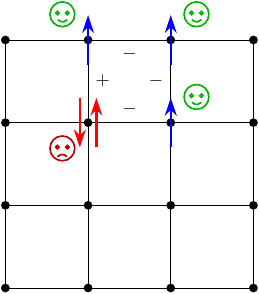}
\caption{Frustrated Ising lattice. We show an example of four spins in which three interaction terms are negative, while on is positive. The blue spins are able to sit in the configuration that minimizes the interactions among them. However, the spin marked in red is {\it frustrated}: both configurations minimize the interaction energy with one partner, but maximize it with the other. This ambiguity leads to a frozen (but with vanishing averaged magnetization) system.}
	\label{fig:Isinglatticefrustrated}
\end{figure}

In this Section we introduce a model that realizes these properties, which is only a slight modification of the one studied in the previous Section. A key modification is to allow the couplings $J_{i_1,\dots ,i_p}$ to take both positive and negative values. This generates a pattern of positive and negative terms in the Hamiltonian that {\it frustrates} the system: the system is not able to minimize all interaction terms simultaneously. This allows the system be frozen in a disordered way, such that the average magnetization vanishes. We show an example of this mechanism in Fig. \ref{fig:Isinglatticefrustrated}.

\subsection{Definition of the model and replica trick}

Consider an ensemble of theories defined by the Hamiltonian \cite{Crisanti:1992sxo}
\begin{equation}
    H=\sum_{i_1 \, < \, \ldots \, < \,i_p}J_{i_1 \ldots i_p}\sigma_{i_1}\cdots \sigma_{i_p} \, ,
\end{equation}
where the spin variables are continuous, $-\infty < \sigma_{i}<\infty$, and are subject to the spherical constraint
\begin{equation}
    \sum_{i} \sigma_{i}^2 = N \, .
\end{equation}
The coupling constants $J_{i_1 \ldots i_p}$ are random variables following a Gaussian distribution of the form
\begin{equation}
    P(J_{i_1 \ldots i_p}) = \frac{1}{J}\sqrt{\frac{N^{p-1}}{p! \cdot \pi}} \exp \left(-\frac{N^{p-1}}{p!}\frac{J_{i_1 \ldots i_p}^2}{J^2}\right) \, .
\end{equation}
This choice of variance ensures a well-defined thermodynamic limit as $N\to\infty$. For $p=2$ and deterministic couplings $J_{ij}=J/N$, the model reduces to the spherical ferromagnet studied earlier. In contrast, for $p\geq 3$ and random couplings of either sign,
the system is intrinsically frustrated: it is impossible to minimize all interaction terms simultaneously

This is known as the $p$-spin spherical model. We now turn to its statistical mechanics.

Because the couplings are random, thermodynamic quantities must be averaged over the disorder. There are two conceptually distinct procedures. In \emph{annealed disorder}, one averages the partition function itself, $\overline{\mathcal{Z}}$, \footnote{The overline indicates an average over the couplings while the $\langle \cdot \rangle$ indicates a thermodynamic average.} and then computes thermodynamic quantities from it. While technically simpler, this procedure does {\bf not} generally yield the correct free energy.

Instead, we must treat the disorder as \emph{quenched}: the free energy is first computed for a
fixed realization of the couplings and only then averaged. The quantity of interest is therefore
\begin{equation}
    \overline{F}
=-\frac{1}{\beta}\,\overline{\log \mathcal Z}\, .
\end{equation}
To evaluate this average, we use the identity
\begin{equation}
\label{log trick}
    \overline{\log(x)}=\frac{d}{dn}\bigg(\overline{x^n} \bigg)\bigg|_{n=0} \, ,
\end{equation}
which is valid upon analytic continuation in $n$.

The idea is then to compute the averaged powers of the partition function, $\overline{\mathcal{Z}^n}$ with $n \in \mathbb{N}$, extrapolate the result to any $n \in \mathbb{R}$, differentiate with respect to $n$, and finally take the $n \rightarrow 0$ limit to find the averaged free energy. This is known as the \textit{replica trick}. 

The partition function of a single copy of the system is
\begin{equation}
    \mathcal{Z}=\int  d\sigma_1 \ldots d\sigma_N \ e^{-\beta H} \delta \!\left(\sum_{i}\sigma_{i}^2 -N\right)\, .
\end{equation}
Powering this expression to the $n$-th power introduces $n$ identical copies (replicas) of the
system, labelled by $a=1,\ldots,n$,
\begin{equation}
    \mathcal{Z}^n=\int \prod_{a=1}^{n}d\sigma_{1}^{(a)} \ldots d\sigma_{N}^{(a)} \ \delta \!\left(\sum_{i}\left(\sigma_{i}^{(a)}\right)^2 -N\right) \exp \left(-\beta H^{(a)} \right)\, .
\end{equation}
Here, $H^{(a)}$ is the Hamiltonian of copy $a$.

We now average over the disorder,
\begin{equation}
    \overline{ \mathcal{Z}^n } = \int dJ_{i_1 \ldots i_p} P(J_{i_1 \ldots i_p}) \mathcal{Z}^n \, .
    \label{eq:disorderaverage}
\end{equation}
Since the couplings are Gaussian, this integral can be performed explicitly. The result is a
replicated theory in which different copies interact with one another,
\begin{equation}
    \overline{\mathcal{Z}^n}  \propto \int d\sigma_{i}^{(a)} \ \delta \!\left(\sum_{i}\left(\sigma_{i}^{(a)}\right)^2 -N\right)\exp\!\left(\frac{p!(\beta J)^2}{4N^{p-1}}\!\sum_{i_1 \, < \, \ldots \, < \, i_p}\!\sum_{a,b}\sigma_{i_1}^{(a)}\cdots \sigma_{i_p}^{(a)}\sigma_{i_1}^{(b)}\cdots \sigma_{i_p}^{(b)}\right)\, .
    \label{eq:averagedZ}
\end{equation}
It should be understood in this expression that there is still a $\delta$ for each value of $a$ and that the measure consists of all $\sigma_{i}^{(a)}$'s.

\newexercise[ex:approximation]{ex:approximation}{Argue that at large $N$, we may apply the following approximation:
$$p! \sum_{1\leq i_1<\ldots < i_p \leq N} = \sum_{i_1,\ldots,i_p=1}^{N} \, .$$}

Using this simplification, the interaction term factorizes as
\begin{equation}
    p! \sum_{1\leq i_1<\ldots < i_p \leq N} \sigma_{i_1}^{(a)}\cdots \sigma_{i_p}^{(a)}\sigma_{i_1}^{(b)}\cdots \sigma_{i_p}^{(b)} = \left(\sum_{i}\sigma_{i}^{(a)}\sigma_{i}^{(b)} \right)^p \, .
\end{equation}

At this stage it is natural to introduce the \emph{overlap} between replicas, which is defined as
\begin{equation}
    \label{eq:Qmatrix}
    Q_{ab}=\frac{1}{N}\sum_{i=1}^{N} \overline{\langle \sigma_{i}^{a}\sigma_{i}^{b} \rangle} \, .
\end{equation}
Moreover, we insert the identity,
\begin{equation}
    \begin{aligned}
        1 &= \int dQ_{ab} \, \delta\!\left(NQ_{ab}-\sum_{i=1}^{N}\sigma_{i}^{(a)}\sigma_{i}^{(b)} \right) \\
        &= \int d\lambda_{ab}\int dQ_{ab} \exp \left(i\lambda_{ab}\left(NQ_{ab}-\sum_{i=1}^{N}\sigma_{i}^{(a)}\sigma_{i}^{(b)}  \right) \right) \, ,
        \label{eq:insert1}
    \end{aligned}
\end{equation}
for each pair of replicas $(a,b)$. Here, $\lambda_{ab}$ plays the role of a Lagrange multiplier. The disorder-averaged partition function becomes
\begin{equation}
    \overline{\mathcal{Z}^n} \propto\!\int \!dQ_{ab}d\lambda_{ab}d\sigma_{i}^{(a)}\delta(Q_{aa}-1)\exp \left[\frac{(\beta J)^2N}{4}\sum_{a,b}(Q_{ab})^p+i\sum_{a,b} \lambda_{ab}\!\left(\!NQ_{ab}-\sum_{i}\sigma_{i}^{(a)}\sigma_{i}^{(b)}\!\right) \right] \, .
\end{equation}
The integrals over the spins are Gaussian and can now be performed, yielding
\begin{equation}
    \overline{\mathcal{Z}^n} \propto\!\int \!dQ_{ab}d\lambda_{ab}\delta(Q_{aa}-1)\exp \left[-N \left(-\frac{(\beta J)^2}{4}\sum_{a,b}(Q_{ab})^p+i\sum_{a,b}\lambda_{ab}Q_{ab}-\frac{1}{2}\mathrm{Tr}\log\!\left(\frac{i \lambda}{\pi} \right)\right)\right] \, .
\end{equation}

In the large-$N$ limit, the saddle-point equation \cite{Bender:1999box} for $\lambda_{ab}$ yields the matrix equation
\begin{equation}
    \lambda=\frac{1}{2i}Q^{-1}
\end{equation}
which, when inserted back into the integral, gives
\begin{equation}
\label{eq:before solving for Q}
    \overline{\mathcal{Z}^n} \propto\!\int \!dQ_{ab}\delta(Q_{aa}-1)\exp \left[-N \left(-\frac{(\beta J)^2}{4}\sum_{a,b}(Q_{ab})^p-\frac{1}{2}\mathrm{Tr}\log Q\right)\right] \, .
\end{equation}

Before solving the saddle-point equations, it is worth pausing to interpret the matrix $Q_{ab}$ defined in \eqref{eq:Qmatrix}. Since different replicas do not interact microscopically, we expect a factorization of the form
\begin{equation}
    \label{factorized Q}
    Q_{ab}=\frac{1}{N}\sum_{i}\overline{\langle \sigma_{i}^{a}\rangle\langle \sigma_{i}^{b}\rangle} \, \, , \, \, a \neq b .
\end{equation}
Moreover, because the system is disordered, the average magnetization vanishes,
\begin{equation}
    \label{average magnetization}
    m^{a}=\frac{1}{N}\sum_{i}\overline{\langle \sigma_{i}^{a} \rangle}=0\, .
\end{equation}
Unlike the ferromagnetic spherical model, there will never be a spontaneous magnetization. Nevertheless, the overlaps $Q_{ab}$ may be non-zero. At high temperatures we expect $Q_{ab}=0$ for $a\neq b$, reflecting the fact that different replicas explore the phase space independently. At low temperatures, this need no longer be true: replicas may become trapped in related regions of configuration space, leading to a non-trivial structure for $Q_{ab}$.

In the large-$N$ limit, the replicated partition function \eqref{eq:before solving for Q} can be written as
\begin{equation}\label{eq:overlineZ}
    \overline{\mathcal{Z}^n} \propto\!\int \!dQ_{ab}\delta(Q_{aa}-1)\exp \left[-N F(Q)\right] \, ,
\end{equation}
with
\begin{equation}\label{eq:F(Q)}
    F(Q) \equiv - \frac{(\beta J)^2}{4}\sum_{a,b}(Q_{ab})^p-\frac{1}{2}\mathrm{Tr}\log Q \, .
\end{equation}
The thermodynamic behaviour of the model is determined by the saddle points of $F(Q)$. Understanding its possible saddle points and their physical meaning is the central goal of what follows.

\subsection{The replica symmetric ansatz}

A natural first step is to assume that all replicas are statistically equivalent. This leads to the replica-symmetric (RS) ansatz \cite{Sherrington:1975zz}
\begin{equation}
\label{replica-symmetric Q}
    Q_{RS}=\begin{pmatrix}
1      & u      & \cdots & u      \\
u      & 1      & \cdots & u      \\
\vdots & \vdots & \ddots & \vdots \\
u      & u      & \cdots & 1
\end{pmatrix} \, ,
\end{equation}
where $-1<u<1$ is the Edwards-Anderson parameter \cite{Edwards:1975aas}.

\newexercise[ex:RSansatz]{ex:RSansatz}{Show that the eigenvalues of the replica-symmetric ansatz $Q_{RS}$ are $\lambda_1=1-u$ of multiplicity $n-1$ and $\lambda_2=1+(n-1)u$ of multiplicity $1$.}

\begin{figure}[t]
    \centering
    \begin{subfigure}[t]{0.48\textwidth}
        \centering
        \includegraphics[width=\textwidth]{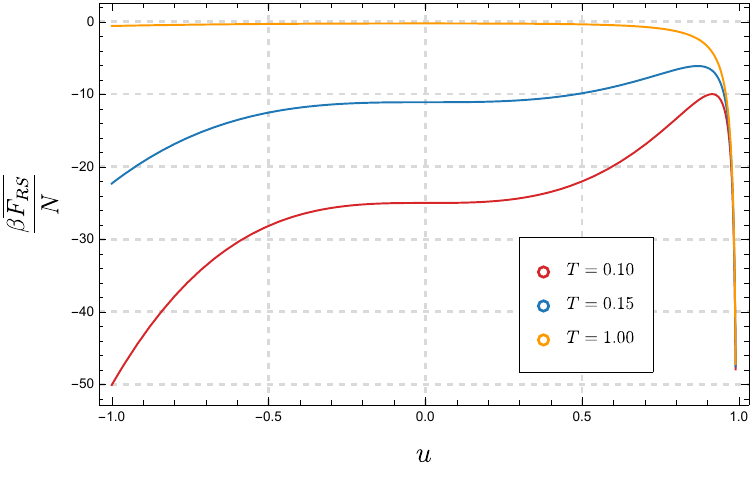}
    \end{subfigure}
    \hspace{0.02\textwidth}
    \begin{subfigure}[t]{0.48\textwidth}
        \centering
        \includegraphics[width=\textwidth]{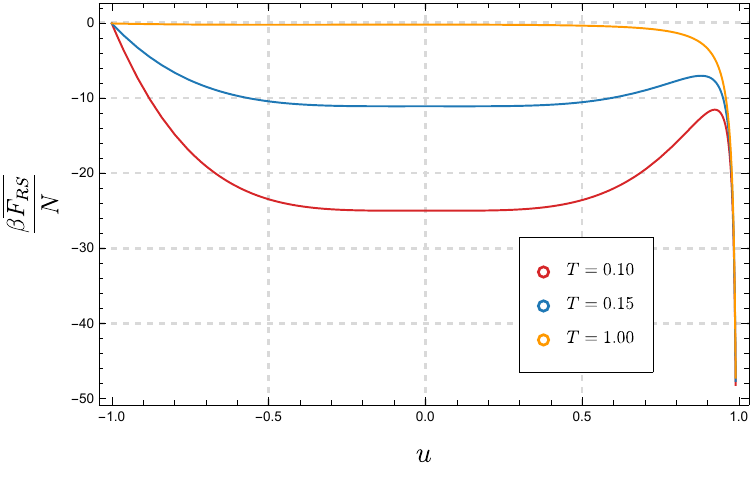}
    \end{subfigure}
    \caption{$\beta \overline{F_{RS}}/N$ as a function of $u$ for various values of $T\equiv(\beta J)^{-1}$, with $p=3$ and $p=4$ respectively.}
    \label{fig:free_energy_spherical_p_spin}
\end{figure}

Using this ansatz, one can simplify \eqref{eq:before solving for Q} using the results of Exercise 3.2 as well as the fact that $\mathrm{Tr}\log=\log \det$ so that
\begin{equation}
    \label{F(Q) simplified}
    F(Q_{RS})=-\frac{(\beta J)^2}{4} (n+n(n-1)u^p)-\frac{1}{2}\log((1-u)^{n-1}(1+(n-1)u)) \, .
\end{equation}
Then, to apply the identity \eqref{log trick}, we first take the $n$-derivative of equation \eqref{eq:overlineZ} with $F(Q)$ replaced by \eqref{F(Q) simplified} (assuming the result extends to $n \in \mathbb{R}$):
\begin{equation}
    \label{n-derivative of overline z}
    \partial_{n}\overline{\mathcal{Z}^n} \propto \int du \, N\left[ \frac{(\beta J)^2}{4}(1+(2n-1)u^p)+\frac{1}{2}\log(1-u)+\frac{1}{2}\frac{u}{1+(n-1) \,u} \right] e^{-NF(Q_{RS})} \, .
\end{equation}
Next, we now take $n \rightarrow 0$. Observing that $F(Q_{RS}) \rightarrow 0$ in this limit, we obtain the following partially off-shell free energy:
\begin{equation}
    \label{eq:RS Free energy}
    \frac{\beta \overline{F_{RS}}}{N}=-\frac{1}{N}\lim_{n \rightarrow 0}\partial_{n}\overline{\mathcal{Z}^n}=-\frac{1}{2} \left[\frac{(\beta J)^2}{2} \left(1-u^p \right)+\frac{u}{1-u}+\log(1-u) \right] \, .
\end{equation}
Extremizing with respect to $u$, one finds
\begin{equation}
    \label{RS free energy extremization}
    \frac{(\beta J)^2}{2}pu^{p-1}-\frac{u}{(1-u)^2} = 0 \, .
\end{equation}
For $p\geq 3$, the solution $u=0$ exists and is stable\footnote{Note that a subtlety with the $n \rightarrow 0$ limit forces us to maximize rather than minimize the free energy. See \cite{Anous:2021eqj} for more details.} at high temperatures. In particular, the variable $u$ quantifies the average correlation of spins across replicas i.e., we do not expect any correlations across the replicas. Its free energy is
\begin{equation}
    \label{u=0 free energy}
    \frac{\beta\overline{F_{RS}}}{N}\Bigg|_{u=0}=-\frac{(\beta J)^2}{4} \, .
\end{equation}
However, as the temperature is lowered, it seems possible to have maximum free energy configurations that support cross-replica correlations (see Fig. \ref{fig:free_energy_spherical_p_spin}).

%\begin{figure}[H]
 %   \centering
  %  \subfigure[$p=3$]{
   %     \includegraphics[width=0.48\textwidth]{plots/plot_free_energy_p=3.pdf}
    %    \label{fig:p3}}
   %  \subfigure[$p=4$]{
     %   \includegraphics[width=0.48\textwidth]{plots/plot_free_energy_p=4.pdf}
    %    \label{fig:p4}}
   % \hfill
  %  \caption{$\frac{\beta \overline{F_{RS}}}{N}$ as a function of $u$}
  %  \label{fig:spherical_p_spin_free_energy}
%\end{figure}

However, this does {\bf not} mean that the model has been solved. For any saddle point to be of minimal energy, deviation in any direction should cause the energy to increase. i.e.,
\begin{equation}
    F_{RS}\left[Q_{ab}+\delta Q_{ab}\right]=\delta Q_{ab}G_{ab|pq}\delta Q_{pq}
\end{equation}
where the matrix $G_{ab,pq}$ needs to have all positive eigenvalues. A detailed stability analysis shows that this condition is not satisfied by the replica-symmetric ansatz \cite{Crisanti:1992sxo}. 

The replica-symmetric solution is therefore not a true minimum of the free energy. To find the physical glassy phase, we must allow for \emph{replica symmetry breaking}.

\subsection{Replica symmetry breaking}
% In the previous lecture, we computed the free energy of the replica symmetric ansatz for the correlation matrix $Q_{ab}$. In particular, the variable $u$ quantified the average correlation of spins across replicas. At high energies, the free energy is minimised for $u=0$, i.e., we do not expect any correlations across the replicas. But, as we cool the glass down, it seems possible to have minimum free energy configurations that support cross-replica correlations. Nevertheless, within the replica symmetric ansatz, these solutions are unstable. 

The simplest way to break the replica symmetry is to partition replicas into smaller blocks of size $m\times m$. Within the blocks, the replicas all have the same overlap: $u$, whereas replicas not within the same block have the overlap $s$. Thus, the structure of the correlation matrix follows the pattern:
\begin{equation}\label{eq:1RSBAnsatz}
\setcounter{MaxMatrixCols}{20}
\setlength{\arraycolsep}{2pt}
\renewcommand{\arraystretch}{0.7}
\setlength{\jot}{0pt}
Q =
  \begin{pmatrix}
   1 & u       & \ldots & u   &   &       &     &   &            &   &   &   &    \\
   u & 1  &     & u  &     &  s & &  &  & & &  s &    &  \ldots   &    s & \\
   \vdots &    &  \ddots    &  \vdots &      &     &   &     &   &   &   & \\
   u & u     & \ldots   & 1   &   &       &     &   &            &   &   &   &  \\
\multicolumn{4}{c}{\raisebox{.5\normalbaselineskip}[0pt][0pt]{$\underbrace{\phantom{ u  u \ldots 1}\hspace*{2\tabcolsep}}$}}  &           \\
\multicolumn{4}{c}{\scriptstyle m}&     &   &            &   &   &   &  &   &\\
   &  &          &     &  1 & u       & \ldots & u      \\
 & & s & & u & 1  &     & u   &  & & & s  & & \ldots & s &  & \\
 & & & &  \vdots &    &  \ddots    &  \vdots &      &     &   &     &  \\
 & & & &  u & u     & \ldots   & 1  &   &        &   & & &   \\
 & & & &\multicolumn{4}{c}{\raisebox{.5\normalbaselineskip}[0pt][0pt]{$\underbrace{\phantom{ u  u    \ldots   1 }\hspace*{2\tabcolsep}}$}} \\
 & & & & \multicolumn{4}{c}{\scriptstyle m}&  & & & \\
  &     &        &     &   &    &   &    & & &  \ddots  & &  & \vdots & \\
   &    &  s    &     &   & s  & & &    & & & \\
   &       &        &     &   &        & & &   & & & \\
   & &\vdots &  &        &   \vdots  &   &   & &     & & &   1 & u       & \ldots & u  \\
   & &      &     &   &   & & &  & &  & &  u & 1 & u  \\
    &     & & & & & & &  & & & &  \vdots &    &  \ddots    &  \vdots \\ 
  &  & s &        &   &   s  &  &  &  & &   \ldots     & &  u   & u   &\ldots   & 1  \\
  & &  & & & & & & & & & & \multicolumn{4}{c}{\raisebox{.5\normalbaselineskip}[0pt][0pt]{$\underbrace{\phantom{ u  u    \ldots   1 }\hspace*{2\tabcolsep}}$}} &\\
  &  &  & & & & & & & & & & \multicolumn{4}{c}{\scriptstyle m}&\\
\multicolumn{20}{c}{\raisebox{.5\normalbaselineskip}[0pt][0pt]{$\underbrace{\phantom{ u u u  u  u u u u   \ldots u  1 1 u u u 1 u u u 1 u u u }\hspace*{5\tabcolsep}}$}} \\
\multicolumn{20}{c}{\scriptstyle n}\\
\noalign{\vspace{-2\normalbaselineskip}}
  \end{pmatrix} \vspace{2\normalbaselineskip}\, ,
\end{equation}
and we choose $m$ such that $m$ divides $n$. This ansatz is referred to as the 1-step replica symmetry breaking ansatz, or $1-$RSB\cite{Parisi:1979ad,Parisi:1979mp,mezard1988spin}. The breaking of the replica symmetry arises due to configurations with $s\neq u$. The `1-step' is due to the fact that there are only two types of correlations between replicas: correlations of replicas in the same block vs correlations between replicas in separate blocks. It follows that, in order to get a 2-step RSB matrix, we build a matrix like \eqref{eq:1RSBAnsatz} but instead we populate the diagonal blocks with 1-RSB matrices. This process can then be repeated to obtain multistep replica symmetric breaking. Such structures follow the mathematics of ultrametric spaces\cite{Parisi:1983dx,mezard1984nature,mezard1988spin}. We will focus only on the 1-RSB ansatz in these notes.

\newexercise[ex:RSBsolution]{ex:RSBsolution}{\begin{enumerate}
    \item Show that, in the $1-$RSB correlation matrix the number of :
    \begin{enumerate}
        \item $1$'s is $n$,
        \item $u$'s is $n(m-1)$,
        \item and $s$'s is $n(n-m)$.
    \end{enumerate}
    \item Show that the eigenvalues of this matrix are 
    \begin{enumerate}
        \item $1-u$ with multiplicity $n(1-\frac{1}{m})$,
        \item $1-u+m(u-s)$ with multiplicity $\frac{n}{m}-1$,
        \item and $1-u+m(u-s)+ns$ with multiplicity $1$.
    \end{enumerate}
\end{enumerate}}

Given the 1-RSB ansatz, we can compute the overlap distribution of the Edwards-Anderson parameter: 
\begin{equation}
    \overline{P(q)}=\frac{m-1}{n-1}\delta(q-u)+\frac{n-m}{n-1}\delta(q-s)\ ,
\end{equation}
where we know that $m$ lies between $1$ and $n$. Since we need to take the limit $n\to 0$, we need to make sense of the above expression in said limit. Strictly speaking, one cannot simultaneously satisfy the limit as well as the condition that $m\leq n$, while keeping $m$ above $1$. Nevertheless, we force the limit in the expression, keeping $m$ constant, to obtain:
\begin{equation}\label{eq:1RSBProbDist}
    \lim_{n\to 0}\overline{P(q)}=(1-m)\delta(q-u)+m\delta(q-s)\ .
\end{equation}
From this expression, we can glean that $m$ has to be a real number such that $0\leq m\leq 1$ for the probabilities to be positive, and henceforth we will treat it as such. Finally, we have our correlation matrix in the $n\to 0$ limit parametrised by $\{u,s,m\}$, whose values we can fix in the usual saddle-point approximation.

%     From the previous lecture, we have, 
% \begin{equation}
%    \overline{\mathcal{Z}^n}  \propto\!\int \!dQ_{ab}\delta(Q_{aa}-1)\exp \left[N \left(\frac{(\beta J)^2}{4}\sum_{a,b}(Q_{ab})^p+\frac{1}{2}\mathrm{Tr}\log Q\right)\right] \, .
% \end{equation}
% This formula was obtained in the saddle point approximation at large $N$. 
Substituting our $1-$RSB ansatz into the averaged partition function \eqref{eq:overlineZ} expression yields:
\begin{equation}
\begin{split}
   \overline{\mathcal{Z}^n}  \propto\ \int_{-1}^{1} du \int_{-1}^{1}ds\ \sum_{m}\exp \left\{N n \left[\frac{\beta^2 J^2}{4}\left(1+u^p(m-1)+s^p(n-m)\right)\right. \right.\\\left.\left.+\frac{1}{2}\left(1-\frac{1}{m}\right)\log{(1-u)}+\frac{1}{2}\left(\frac{1}{m}-\frac{1}{n}\right)\log\left(1-u+m(u-s)\right)\right.\right. \\ \left.\left. +\frac{1}{2n}\log\left(1-u+m(u-s)+ns\right)\right]\right\} \, .
\end{split}
\end{equation}

Since we converted $m$ to a real number between $0$ and $1$, such that it makes sense in the $n\to 0$ limit, we also changed the sum in the above expression to an integral from $0$ to $1$.

Given this explicit expression, we can compute the free energy:
\begin{equation}
\begin{split}
    -2\beta\frac{\overline{F}}{N}= & \frac{\beta^2J^2}{2}\left[1-u^p+m(u^p-s^p)\right] +\frac{m-1}{m}\log(1-u)  \\ & +\frac{1}{m}\log\left[m(u-s)+(1-u)\right]+\frac{s}{m(u-s)+(1-u)}\ .
\end{split}
\end{equation}
Now, similarly to the replica-symmetric case, we want to find the minima of the free energy by varying the parameters $u$, $s$ and $m$. With that in mind, we set the variations of the free energy with respect to these parameters to zero and study the corresponding solutions. Let's begin with varying with respect to $s$, i.e. overlaps between replicas not belonging to the same $m\times m$ block:
\begin{equation}
    \frac{\partial \overline{F}}{\partial s}= 0 \implies m s \left[\frac{1}{\left(1-u+m(u-s)\right)^2}-\frac{\beta^2J^2}{2}p s^{p-2}\right]=0
\end{equation}
As can be gleaned from the above equation, for $p>0$, $s=0$ is a solution at any temperature. This is expected for disordered systems, since we expect that randomly drawn states are uncorrelated. Hence, we set $s=0$ moving forward. 

\begin{figure}[t]
    \centering
    \includegraphics[width=0.7\linewidth]{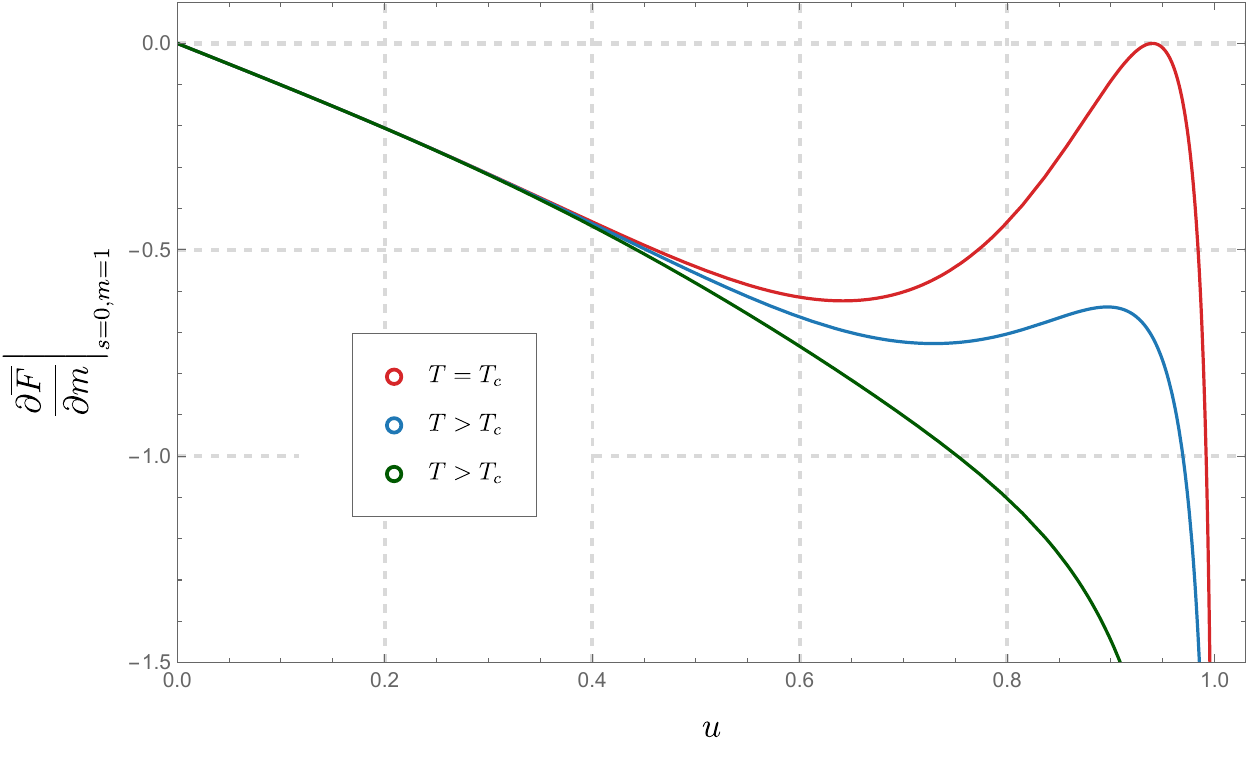}
    \caption{Non-trivial zeros for the EOM for the 1-RSB ansatz.}
    \label{fig:1RSBPlot}
\end{figure}

Varying the free energy with respect to $u$ and $m$ gives us the following two equations:
\begin{equation}
\begin{aligned}
    \frac{\partial \overline{F}}{\partial u}=0 &\quad\implies\quad (1-m)\left(\frac{\beta^2J^2}{2}pu^{p-1}-\frac{u}{(1-u)\left[(m-1)u+1\right]}\right)=0 \ ,\\
   \frac{\partial \overline{F}}{\partial m}=0 &\quad\implies\quad \frac{\beta^2J^2}{2}u^{p-1}+\frac{u}{m+m(m-1)u}+\frac{1}{m^2}\log\left[\frac{1-u}{1+(m-1)u}\right]=0\ .
   \label{eq:Eqsuandmstatic}
\end{aligned}
\end{equation}
These equations can now be studied numerically to understand the behaviour of the phase transition. One finds that for high temperatures, i.e. small $\beta$, $u=0$ is the only solution. If $u\neq 0$, then one can satisfy the 1st of the above equations by setting $ m=1$. Then we study the numerical behaviour of the LHS of the second equation for $m=1$, viz.
\begin{equation}
    \frac{\beta^2J^2}{2}u^{p-1}+u+\log\left[1-u\right]\ .
\end{equation}
We can plot this quantity for various values of $\beta J$ and find that, for temperatures above a certain value $T_c$, this expression is zero only when $u=0$, but for temperatures below $T_c$, there exist nontrivial solutions for $u$ (see Fig. \ref{fig:1RSBPlot}). Unlike the replica-symmetric case, these solutions are stable and provide us with non-trivial replica overlaps under the critical temperature. 

Let's try to understand what the replica-symmetry-breaking is telling us about the glassy phase. Unlike usual phase transitions where thermal fluctuations compete with the tendency of a system to order itself, we have here a phase transition that is \emph{driven by disorder}. The presence of random couplings allows for frustration at low energies. This does not allow the system to settle down into a unique, simple, ordered state. Instead, the system can settle into any of a large number of low-energy configurations. The glassiness emerges specifically as a result of quenched disorder, i.e., the random couplings stay constant over the thermal averaging. Such frozen couplings signal a lack of ergodicity for experimental time-scales: the system is trapped in a particular corner of the configuration space. This is the structure that our RSB ansatz is exploring. It is a conditioned thermal probability distribution for fixed given values of couplings.   

In our simple $1-$RSB picture, the block structure emerges from several replicas settling into a thermodynamic `pure state': a single valley in the landscape of free energy. The pure state only explores small thermal fluctuations within the valley but never shifts into another valley on experimental time scales. All replicas belonging to the same pure state find themselves in a single block due to high cross-replica correlations. Whereas replicas belonging to different pure states have almost no overlap, as they belong to completely different valleys. This structure is obvious from the overlap distribution $P(q)$, which encodes the two-level structure of interblock and intrablock correlations. Hence, the RSB structure captures the physics of the quenched disorder as a distribution over such pure states, and is not simply a mathematical trick\footnote{See chapter 3 of \cite{mezard1988spin} for an extended physical interpretation.}.

Given our unique picture of a glassy phase transition, the next question is obvious. Can an experiment delineate for us this phase transition through sharp changes in measured susceptibilities, as in the case of magnetism? Unlike magnetisation, which is a macroscopic observable, our `order-parameter' for the glass phase transition, $u$, lives in replica space and is not easily realisable as an experimental observable. The transition into a glassy state materialises instead as a divergence in the system's relaxation time scales\cite{kirkpatrick1987p,cugliandolo1993analytical}. Such an observable can only be inferred through the study of dynamics, for which a purely static analysis like ours is too limiting. Hence, we will move on to understanding the dynamical picture of the spin-glass, so that we may be able to say more about sharp observables for glassy phase transitions.

\section{Quantum dynamics}\label{sec:dynamics}
We consider a quantum version of the spherical $p$-spin model, defined by the Euclidean action
\begin{equation}
    S_E=\int_0^\beta d\tau\left[\frac{M}{2}\sum_{i=1}^N\Dot{\sigma}_i(\tau)\Dot{\sigma}_i(\tau)+\sum_{i_1<...<i_p}J_{i_1...i_p}\sigma_{i_1}(\tau)...\sigma_{i_p}(\tau)\right]~,
\end{equation}
with the spherical constraint
\begin{equation}
    \sum_{i=1}^N\sigma_i(\tau)\sigma_i(\tau)=N~.
\end{equation}
The parameter $M$ controls the strength of quantum fluctuations, with $M\to\infty$ being the classical limit.

\subsection{Euclidean formulation}
As in the static analysis, we will use the replica trick to write a large-$N$ effective action for the model. Proceeding as in Section \ref{sec:sphericalpspin}, but now including the kinetic term, one finds the replicated partition function
\begin{equation}
\begin{split}
	&\overline{Z^n}=\int DQ_{ab}D\lambda_{ab} Dz^a \;\text{det}^{-N/2}\left[-\frac{1}{\pi}\delta_{ab}\delta(\tau-\tau')\left(\frac{M}{2}\partial_{\tau'}^2+i\, z^a(\tau)\right)+\frac{i}{\pi}\lambda_{ab}(\tau,\tau')\right]\\
	&\quad \times \exp\left\lbrace-N\,i\sum_{a=1}^n\int_0^\beta d\tau\,  z^a(\tau)+N\int_0^\beta \int_0^\beta d\tau\,d\tau'\,\left[i\lambda_{ab}(\tau,\tau')Q_{ab}(\tau,\tau')+\frac{J^2 }{4}\sum_{a,b=1}^n Q_{ab}(\tau,\tau')^p\right]\right\rbrace~,
\end{split}
\end{equation}
from which we read the replicated effective action
\begin{multline}
	\frac{S_{\rm eff}}{N}=\frac{1}{2}\text{Tr}\log \left[-\frac{1}{\pi}\delta_{ab}\delta(\tau-\tau')\left(\frac{M}{2}\partial_{\tau'}^2+i\, z^a(\tau)\right)+\frac{i}{\pi}\lambda_{ab}(\tau,\tau')\right]\\+i\sum_{a=1}^n\int_0^\beta d\tau\,  z^a(\tau)-\int_0^\beta \int_0^\beta d\tau\,d\tau'\,\left[i\lambda_{ab}(\tau,\tau')Q_{ab}(\tau,\tau')+\frac{J^2 }{4}\sum_{a,b=1}^n Q_{ab}(\tau,\tau')^p\right]~.
\end{multline}

The bilocal field $Q_{ab}$ is defined analogously to the static one,
\begin{equation}
    Q_{ab}(\tau,\tau')\equiv \frac{1}{N}\sum_{i=1}^N\langle \sigma_i^a(\tau)\sigma_i^b(\tau)\rangle~.
\end{equation}

Again, we can eliminate the dependence on the matrix $\lambda$ by solving its equation of motion,
\begin{equation}
	\left(\mathbf{M}+i\boldsymbol{\lambda}\right)^{-1}=\frac{1}{\pi}\mathbf{Q} \,,
\end{equation}
where we have defined the matrix $\mathbf{M}$ in replica space and imaginary time: 
\begin{equation}
	{M}_{ab}(\tau,\tau')=-\delta_{ab}\delta(\tau-\tau')\left(\frac{M}{2}\partial_\tau^2+i\, z^a(\tau)\right)~.
\end{equation}
This allows us to write the large-$N$ effective action
\begin{equation}\label{eq:effaction}
\begin{aligned}
	\frac{S_{\rm eff}}{N} &=-\frac{1}{2}\text{Tr}\log \left[Q_{ab}(\tau,\tau')\right]-i\sum_{a=1}^n\int_0^\beta d\tau\,  z^a(\tau)\left(Q_{aa}(\tau,\tau)-1\right)\\
	&\quad\, -\sum_{a,b=1}^n\int_0^\beta \int_0^\beta d\tau\,d\tau'\,\left[\delta_{ab}\,\delta(\tau-\tau')\frac{M}{2}\partial_\tau^2Q_{ab}(\tau,\tau')+\frac{J^2 }{4} Q_{ab}(\tau,\tau')^p\right]~.
\end{aligned}
\end{equation}
Since the effective action is proportional to $N$, the saddle-point approximation is exact in the $N\to\infty$ limit.

In order to derive the equations of motion from this action, we first need to write the dynamical version of the 1-RSB ansatz \eqref{eq:1RSBAnsatz}. We begin by noticing that the matrix elements $Q_{a\neq b}$ are time-independent \cite{BrayMoore_1980}
\begin{equation}
	Q_{a\neq b}(\tau,\tau')=\frac{1}{N}\sum_{i=1}^N\overline{\langle \sigma_i^a(\tau)\sigma_i^b(\tau')\rangle}=\frac{1}{N}\sum_{i=1}^N\overline{\langle \sigma_i^a(\tau)\rangle\langle\sigma_i^b(\tau')\rangle}~,
\end{equation}
Since the replica index is not the same, the thermodynamic averages split up into single replica expectation values --- but the one point functions are time independent (although crucially nonzero). This means only the diagonal elements have any time dependence. The late-Euclidean time limit of this diagonal element is the Edwards-Anderson order parameter $u$ we have already encountered in the static analysis \cite{Edwards:1975aas}.

Therefore, the same structure for the RSB ansatz as in \eqref{eq:1RSBAnsatz} works in this case, with the only difference being that the diagonal elements become time-dependent. We can write an ansatz of the form
\begin{equation}
\label{eq:1rsb}
	\mathbf{Q}=\begin{pmatrix} 
	q(\tau,\tau') & u & u &   &  & &\\
	u & q(\tau,\tau') &u &  &s & &\cdots\\
	u & u & q(\tau,\tau') & & & &\\
	  &  &  & q(\tau,\tau') &  u& u &\\
	  &  s&  & u & q(\tau,\tau') &u &\\
	  &  &  & u & u & q(\tau,\tau') &\\
	  & \vdots & & & & & \ddots &~,
	\end{pmatrix}
\end{equation}
where the $n\times n$ matrix is made up of a set of $m\times m$ blocks along the diagonal, with $m$ to be determined.

Using the results of Exercise \ref{ex:RSBsolution}, the effective action \eqref{eq:effaction} can be written in terms of the RSB parameters as
\begin{equation}
\begin{aligned}\label{eq:1rsbeffaction}
	\frac{S_{\rm eff}}{N}=&-\frac{n}{2}\left(1-\frac{1}{m}\right)\text{Tr}\log \left[q(\tau,\tau')-u\right]-\frac{1}{2}\left(\frac{n}{m}-1\right)\text{Tr}\log \left[q(\tau,\tau')-u +m(u-s)\right]\\&-\frac{1}{2}\text{Tr}\log\left[ q(\tau,\tau')-u+m(u-s)+ns\right]-in\int_0^\beta d\tau\,  z(\tau)\left(q(\tau,\tau)-1\right)\\
	&-n\int_0^\beta \int_0^\beta d\tau\,d\tau'\,\left\lbrace\delta(\tau-\tau')\frac{M}{2}\partial_\tau^2q(\tau,\tau')+\frac{J^2 }{4} \left[q(\tau,\tau')^p+(m-1)u^p+(n-m)s^p\right]\right\rbrace~.
\end{aligned}
\end{equation}

This form of the action is useful to derive the equations of motion for the off-diagonal parameters. Varying with respect to $s$ one obtains
\begin{equation}
\begin{aligned}
\text{EOM}_s: \quad\;\; 0&=\frac{pJ^2}{4}\int_0^\beta d\tau''\left[q(\tau,\tau'')^{p-1}s+s^{p-1}q(\tau'',\tau')+(m-1)\left(s\, u^{p-1}+s^{p-1}u\right)+(n-2m)s^p\right]~.
\label{eq:eomvarys}
\end{aligned}
\end{equation}
We see that this is proportional to $s$, so it is safe to set $s$ to zero. The equation of motion for varying $u$ takes a similar form (for $s=0$):
\begin{equation}
\begin{aligned}
\label{eq:eomvaryu}
\text{EOM}_u: \qquad 0&=\frac{pJ^2}{4}\int_0^\beta d\tau''\left[q(\tau,\tau'')^{p-1}u+u^{p-1}q(\tau'',\tau')+(m-2)u^p\right]~.
\end{aligned}
\end{equation}
From this, one might also conclude that $u=0$ by similar logic as above, and indeed, this solution is valid at high temperatures. However, unlike the case with $s$, at low temperatures a new solution with $u\neq0$ appears and dominates the thermodynamics. So we will allow for generic values of $u$ for now.

It is convenient to write the effective action in frequency space. In the subspace $s=0$, the action \eqref{eq:1rsbeffaction} becomes
\begin{equation}
\begin{aligned}
	\frac{S_{\rm eff}}{nN}=&-\frac{1}{2}\sum_{k\neq 0}\log \frac{\hat{q}(k)}{\beta}-\frac{1}{2}\left(1-\frac{1}{m}\right)\log \left[\frac{\hat{q}(0)}{\beta}-  u\right]-\frac{1}{2m}\log \left[\frac{\hat{q}(0)}{\beta}+ (m-1) u\right]\\
	&+i\hat{z}(0)\left(1-\sum_{k=-\infty}^\infty \frac{\hat{q}(k)}{\beta}\right)+\frac{M}{2}\sum_{k=-\infty}^\infty \left(\frac{2\pi k}{\beta}\right)^2\hat{q}(k)\\
	&-\frac{\beta^2J^2 }{4}\left\{(m-1)u^p+\sum_{k_1,\dots,k_{p}}\left[\prod_{i=1}^{p}\frac{\hat{q}(l_i)}{\beta}\right]\,\delta_{k_1+\dots+k_p,0}\right\}~,
    \label{eq:Sfinal}
\end{aligned}
\end{equation}
where we use the conventions for the Fourier transform as
\begin{equation}
    q(\tau,\tau')=\frac{1}{\beta}\sum_{k=-\infty}^\infty e^{\frac{2\pi i k}{\beta}(\tau-\tau')}\hat{q}(k),
\end{equation}
where time-translation symmetry simplifies the saddle-point solutions to be functions only of the time-difference $\tau-\tau'$.

We notice that $u$ and $m$ only couple to the zero mode of $q(\tau,\tau')$. Note moreover that we need only retain the zero mode of $z(\tau)$ in order to impose the spherical constraint:
\begin{equation}
 	\hat{z}(0) \equiv \int_0^\beta d\tau\, z(\tau)~.
 \end{equation}

Varying the action \eqref{eq:Sfinal} with respect to $\hat{z}(0)$, $u$ and $m$ gives, respectively,
\begin{align}
\text{EOM}_z:\qquad & \frac{1}{\beta}\sum_{k=-\infty}^\infty \hat{q}(k) = 1 \label{eq:eomzq}\\
\text{EOM}_u:\qquad{} &\frac{m-1}{2}\left(\frac{\beta^2J^2}{2}p\,u^{p-1}-\frac{u}{\left(\frac{\hat{q}(0)}{\beta}-u\right)\left(\frac{\hat{q}(0)}{\beta}+(m-1)u\right)}\right)=0 \label{eq:eomuq}\\
\text{EOM}_m:\qquad &\frac{\beta^2J^2}{2}u^p+\frac{1}{m}\frac{u}{\frac{\hat{q}(0)}{\beta}+(m-1)u}+\frac{1}{m^2}\log\left[\frac{\frac{\hat{q}(0)}{\beta}-u}{\frac{\hat{q}(0)}{\beta}+(m-1)u}\right]=0 \label{eq:eommq}
\end{align}
Equation \eqref{eq:eomzq} is nothing but the spherical constraint in momentum space. Equations \eqref{eq:eomuq}, \eqref{eq:eommq} are the standard spin glass equations \eqref{eq:Eqsuandmstatic}.

The equation of motion for $\hat{q}(k)$ is a bit more involved, but a remarkable simplification happens if we adopt the following definitions \cite{Cugliandolo_2001}. First, we define the self-energy $\Lambda(\tau,\tau')$ as
\begin{equation}
    \Lambda(\tau,\tau')\equiv \frac{p}{2}q(\tau,\tau')^{p-1}\, .
    \label{eq:Lambdadef}
\end{equation}
It is also very convenient to subtract off the constant spin glass parameters by defining
\begin{equation}
\label{eq:subtractSGparams}
\begin{split}
   q_r(\tau,\tau') &= q(\tau,\tau') - u \,,\qquad \Lambda_r(\tau,\tau') = \Lambda(\tau,\tau') - \frac{p}{2}\, u^{p-1} \,,\\
   \hat{q}_r(k) &= \hat{q}(k) - \beta u \delta_{k,0} \,,\qquad\;
   \hat{\Lambda}_r(k) = \hat{\Lambda}(k) - \frac{p}{2} \beta u^{p-1} \delta_{k,0}
   \end{split}
\end{equation}
with boundary conditions $q_r(\tau,\tau) = 1-u$ and $\Lambda_r(\tau,\tau) = \frac{p}{2} (1-u^{p-1})$.
We provide the final result here, and leave the details of the derivation as an exercise:
\begin{equation}
\label{eq:eomqr}
\boxed{ \;\;\text{EOM}_q:\qquad 	\frac{1}{\hat{q}_r(k)}-\frac{1}{\hat{q}_r(0)}= M \left(\frac{2\pi k}{\beta}\right)^2- J^2 \, (\hat{\Lambda}_r(k)-\hat{\Lambda}_r(0)  ) \, ,\;
 }\end{equation}
which is valid both in the paramagnetic phase ($u=0$) and in the spin glass phase ($u\neq 0$). Of course, for $k=0$ this equation is empty. Moreover, one is still required to impose the spherical constraint \eqref{eq:eomzq}.

Finally, in this parametrization, equations \eqref{eq:eomuq} and \eqref{eq:eommq} become:
\begin{equation}
\label{eq:eomumqr}
\boxed{
\begin{aligned}
 \text{EOM}_u: \qquad 0&=\frac{m-1}{2}\left[\frac{(\beta J)^2}{2}p\,u^{p-1}-\frac{u}{\frac{\hat{q}_r(0)}{\beta}\left(\frac{\hat{q}_r(0)}{\beta}+mu\right)}\right] \\
\;\;\;\text{EOM}_m: \qquad 0&=\frac{(\beta J)^2}{2}u^p+\frac{1}{m}\frac{u}{\frac{\hat{q}_r(0)}{\beta}+mu}+\frac{1}{m^2}\log\left[\frac{\frac{\hat{q}_r(0)}{\beta}}{\frac{\hat{q}_r(0)}{\beta}+mu}\right]\;
\end{aligned}
}
\end{equation}

The set of equations \eqref{eq:eomzq},\eqref{eq:eomqr},\eqref{eq:eomumqr} can be solved numerically using an iterative algorithm similar to the one initially proposed for the SYK model \cite{Maldacena:2016hyu}.

\newexercise[ex:eomqr]{ex:eomqr}{Obtain the equation of motion for $\hat{q}_r(k)$, given in \eqref{eq:eomqr}.}

\begin{figure}[t]
    \centering
    \includegraphics[width=0.6\textwidth]{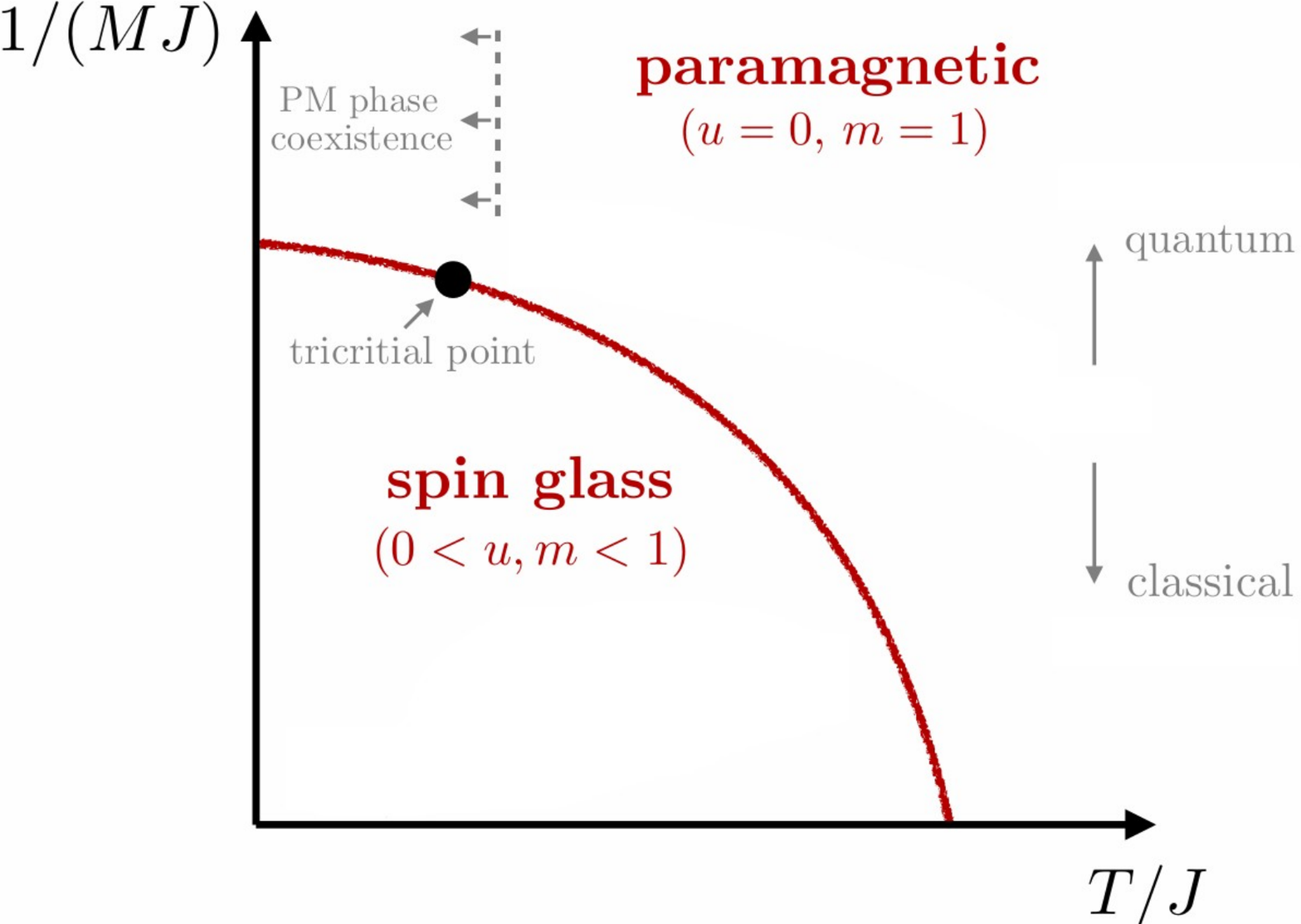}
    \caption{Cartoon of the phase diagram of the $p$-spin spherical model. The basic parameters are the temperature $T/J$ and $1/(MJ)$. We show the rough shape of the spin glass transition, and the associated values of the order parameters $u$ and $m$. Figure adapted from \cite{Anous:2021eqj}.}
    \label{fig:phasediagram}
\end{figure}

Solving numerically the equations of motion, we can obtain the phase diagram of the model. A schematic picture of the phase diagram is shown in Fig. \ref{fig:phasediagram}. The structure of the phase diagram shows two different regions:  a paramagnetic region (where $m=1$ and $u=0$) and a spin glass region (where $0<u<1$ and $m<1$), depending on the value of $MJ$.
\begin{itemize}
    \item For temperatures $T\gtrapprox 0.19 J$ (for $p=3$), the two solutions are separated by a second order phase transition as $MJ$ is tuned. The crossover between the phases is characterized by a continuous (discontinuous) change in $m$ ($u$) across the phase transition line.
    \item For temperatures $T\lessapprox 0.19 J$, the transition becomes first order and the spin glass parameters change discontinuously\footnote{Actually, a third paramagnetic solution exists in this region, but it turns out to be unphysical. See \cite{Anous:2021eqj} for a detailed analysis of the three solutions.}.
\end{itemize}

\begin{figure}[t]
    \centering
    \begin{subfigure}[t]{0.48\textwidth}
        \centering
        \includegraphics[width=\textwidth]{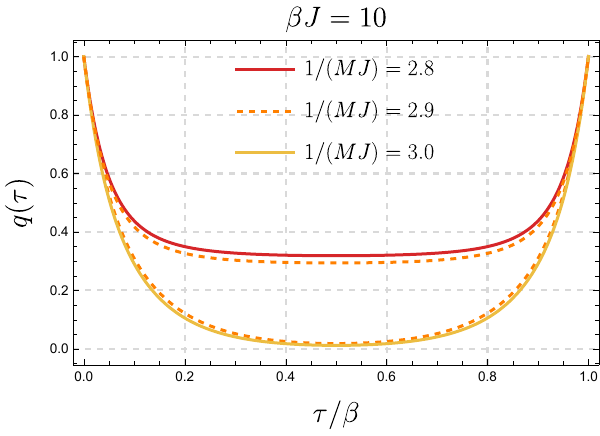}
    \end{subfigure}
    \hspace{0.02\textwidth}
    \begin{subfigure}[t]{0.48\textwidth}
        \centering
        \includegraphics[width=\textwidth]{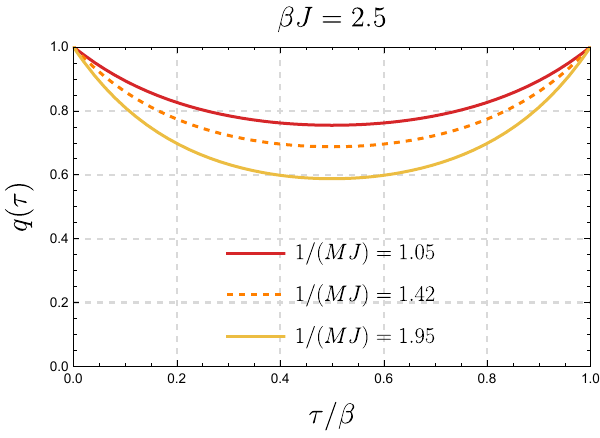}
    \end{subfigure}
    \caption{Reproduction of Fig. 4 from \cite{Anous:2021eqj}. We show the two-point functions at the transition. On the left (low temperatures) the transition is first-order and two solutions coexist near the transition. At higher temperatures, the transition is second order.}
    \label{fig:2pfunctions}
\end{figure}

The behavior of the Euclidean two-point functions is different in each of the two phases, as we show in Fig. \ref{fig:2pfunctions}. Therefore, if we had access to local correlation functions in Euclidean time, we would be able to diagnose the spin glass transition by studying these correlations as a function of the temperature. However, we only have the ability to perform Lorentzian experiments. For this reason, we need to determine how to perform the analytic continuation and define properly the problem in real time.

\subsection{Analytic continuation and real-time correlators}

In order to study real-time dynamics, we first recall how Euclidean thermal correlation functions admit a representation that allows for analytic continuation. Let $\mathcal{O}$ be an operator acting on the Hilbert space $\mathcal{H}$. Its Euclidean two-point function is defined by its thermal expectation value, \textit{i.e.}
\begin{equation}\label{eq:correlator}
	G_{\mathcal{O}\mathcal{O}}(\tau,0)=\Tr_{\mathcal{H}}[e^{-\beta H}\mathcal{O}(\tau)\mathcal{O}(0)],
\end{equation}
where the trace is taken over the Hilbert space $\mathcal{H}$ and $0<\tau<\beta$. Thermal equilibrium implies the KMS condition \cite{Cohen1982}, which for bosonic operators enforces periodicity in imaginary time, that is $G_{\mathcal{OO}}(\tau+\beta)=G_\mathcal{O}(\tau)$. This operator $\mathcal{O}(\tau)$ is the time evolution of $\mathcal{O}(0)$ in Euclidean signature,
\begin{equation}
	\mathcal{O}(\tau)\equiv e^{H \tau}\mathcal{O}(0)e^{-H\tau}.
\end{equation}
After inserting a pair of complete set of states, the correlator in \eqref{eq:correlator} can be written as
\begin{equation}
	G_{\mathcal{O}\mathcal{O}}(\tau,0)=\displaystyle \sum_{n,m} e^{-\beta E_n}e^{(E_n-E_m)\tau}|\bra{n}\mathcal{O}(0)\ket{m}|^2.
\end{equation}
This expression makes manifest that the correlator is controlled by transitions between energy states. We now Fourier transform to Matsubara frequencies \cite{Altland_Simons_2010}, defining $\omega_k=2\pi k/\beta$,
\begin{equation}
	 \hat{G}_{\mathcal{O}\mathcal{O}}(\omega_k)=\int_0^\beta d\tau e^{i\omega\tau}G_{\mathcal{O}\mathcal{O}}(\tau) =\displaystyle \sum_{n,m} e^{-\beta E_n}(1-e^{\beta(E_n-E_m)})\frac{|\bra{n}\mathcal{O}(0)\ket{m}|^2}{-i\omega_k -(E_n-E_m)}\, .
\end{equation}
It is convenient to introduce the density of states $\rho_{\mathcal{O}\mathcal{O}}(\omega)$
\begin{equation}
	\rho_{\mathcal{O}\mathcal{O}}(\omega)=2\pi (1-e^{\beta \omega})\displaystyle\sum_{m,n} e^{-\beta E_n}\delta(\omega-(E_n-E_m))|\bra{n}\mathcal{O}(0)\ket{m}|^2\, .
\end{equation}
This object encodes the dynamical information of the operator $\mathcal{O}$, in terms of which the Matsubara correlator admits the representation
\begin{equation}
	\hat{G}_{\mathcal{O}\mathcal{O}}(\omega_k)=\int_{-\infty}^\infty \frac{d\omega'}{2\pi}\frac{\rho_{\mathcal{O}\mathcal{O}}(\omega')}{-i\omega_k-\omega'}.
\end{equation}
This expression makes the analytic structure very clear: the Matsubara correlator is obtained after evaluating a complex function with  poles on the real axis at discrete imaginary frequencies.

To recover the Euclidean two-point function in time domain we just need to inverse Fourier transform the previous expression,
\begin{equation}
	G_{\mathcal{O}\mathcal{O}}(\tau)=\frac{1}{\beta}\displaystyle \sum_{k}e^{i\omega_k \tau}\hat{G}_{\mathcal{O}\mathcal{O}}(\omega_k)=\frac{1}{\beta}\int_{-\infty}^\infty \frac{d\omega'}{2\pi}\displaystyle \sum_{k}e^{i\omega_k \tau}\frac{\rho_{\mathcal{O}\mathcal{O}}(\omega')}{-i\omega_k-\omega'}.
\end{equation}
We first focus on the sum over $\omega_k$ and, as $\rho_{\mathcal{OO}}(\omega')$ just depends on $\omega'$, the term that we need to compute reduces to 
\begin{equation}
	\frac{1}{\beta}\displaystyle\sum_{k=-\infty}^\infty \frac{e^{i\omega_k \tau}}{-i\omega_k-\omega'}.
\end{equation}
As an intermediate step, introducing $\mu_k=\frac{\beta \omega_k}{2\pi}$, this becomes
\begin{equation}
	\frac{1}{\beta}\displaystyle\sum_{k=-\infty}^\infty \frac{e^{i\omega_k \tau}}{-i\omega_k-\omega'}=\frac{1}{\beta}\displaystyle\sum_{k=-\infty}^\infty \frac{e^{2\pi i\mu_k \tau/\beta}}{-i\frac{2\pi \mu_k}{\beta} -\omega'}.
\end{equation}
Applying the Poisson summation formula \cite{bachmann_2012_fourier}, the Matsubara sum can be rewritten as a contour integral,
\begin{equation}
    \frac{1}{\beta}\displaystyle\sum_{k=-\infty}^\infty \frac{e^{2\pi i\mu_k \tau/\beta}}{-\frac{2\pi i \mu_k}{\beta} -\omega'}=\frac{1}{\beta}\int_{-\infty}^\infty d\mu \sum_{n=-\infty}^\infty e^{2\pi i \mu n}\frac{e^{2\pi i\mu \tau/\beta}}{-i\frac{2\pi \mu}{\beta} -\omega'}=\int_{-\infty}^\infty \frac{d\omega}{2\pi} \sum_{n=-\infty}^\infty e^{i\omega \beta n}\frac{e^{i\omega \tau}}{-i\omega-\omega'}
\end{equation}
In the last step we have done the change $\mu\rightarrow \beta \omega/(2\pi)$. This integral can now be evaluated by closing the contour with a semi-circumference in the complex $\omega$-plane and to apply the residue theorem. As we can see, the integrand has a simple pole located at $\omega=i\omega'$ and the exponential factor is $e^{i\omega(\tau+\beta n)}$. We will assume that $\omega'$ is positive. The choice of contour depends on the sign of $\tau+\beta n$ and, as $0\leq \tau <\beta$, one distinguishes the following two cases:
\begin{itemize}
    \item For $n\geq 0$, the argument of the exponential is positive and the contour has to be closed in the upper half-plane to ensure convergence. The pole $\omega=i\omega'$ lies inside the contour and has a contribution in the integral.
    \item For $n<0$, $\tau+\beta n$ is negative and the contour is closed in the lower half-plane. Here, no pole is enclosed in the region so it does not contribute to the integral.
\end{itemize}

For this reason, only the terms with $n\geq 0$ survive and the previous sum reduces to
\begin{equation}
	\frac{1}{\beta}\displaystyle\sum_{k=-\infty}^\infty \frac{e^{i\omega_k \tau}}{-i\omega_k-\omega}=\displaystyle\sum_{n=0}^\infty e^{-\omega'(\tau+n\beta)}=\frac{e^{-\omega'\tau}}{1-e^{-\beta \omega'}}.
\end{equation}
In the second equality, we have taken into account that the sum is a geometric series. As a result, the Euclidean correlator admits the representation
\begin{equation}\label{eq: green function}
	G_{\mathcal{O}\mathcal{O}}(\tau)=\int_{-\infty}^\infty \frac{d\omega}{2\pi}\frac{e^{-\omega \tau}}{1-e^{-\beta \omega}}\rho_{\mathcal{O}\mathcal{O}}(\omega)\, .
\end{equation}
Notice that this expression is not written in terms of Matsubara frequencies, but the integral is performed over a continuous frequency variable, which lets us perform an analytic continuation of the thermal correlation function in imaginary time into correlation functions in real time.

The passage to Lorentzian signature is achieved by analytic continuation $\tau\rightarrow it$. Different continuations correspond to different operator orderings. We first define
\begin{equation}
	\begin{split}
		G^>_{\mathcal{O}\mathcal{O}}(t)=G(\tau=+it+\epsilon),\\
		G^<_{\mathcal{O}\mathcal{O}}(t)=G(\tau=-it+\epsilon).
	\end{split}
\end{equation}
In the real world, it is not possible to measure correlation functions, but just commutators, which encode information of the order of two events. For this reason, from now on, we will work with the retarded correlation function, defined as
\begin{equation}
	i G_{\mathcal{O}\mathcal{O}}^R(t)=\Tr_{\mathcal{H}}\{e^{-\beta H}\hat{T}[\mathcal{O}(t),\mathcal{O}(0)]\}=\Theta(t)[G^>_{\mathcal{O}\mathcal{O}}(t)-G^<_{\mathcal{O}\mathcal{O}}(t)],
\end{equation}
and its Fourier transform
\begin{equation}
	i G_{\mathcal{O}\mathcal{O}}^R(\omega)= -i \int \frac{d\omega'}{2\pi}\frac{\rho(\omega')}{\omega+\omega'+i\epsilon}.
\end{equation}
Note that these Lorentzian time correlators still contain information about the temperature, both in the $\beta$ appearing in the denominator of \eqref{eq: green function} and in the expression of the density of states.

\begin{figure}[t]
    \begin{subfigure}{0.5\textwidth}
        \centering
        \includegraphics[width=0.8\linewidth]{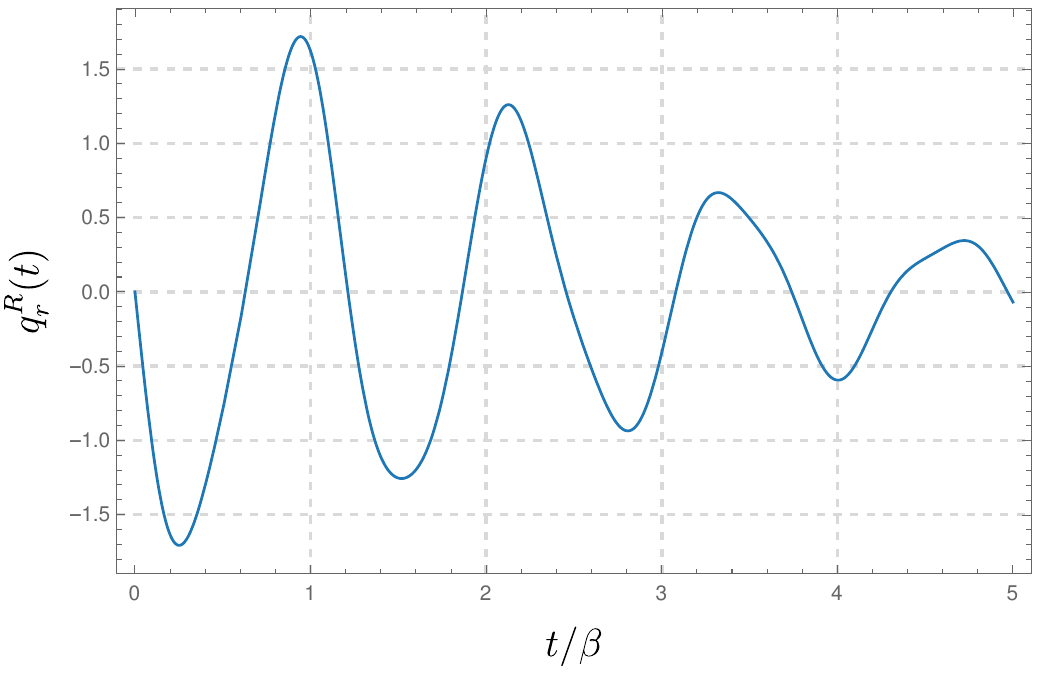}
        \caption{Small values of $MJ$.}
    \end{subfigure}
    \begin{subfigure}{0.5\linewidth}
        \centering
        \includegraphics[width=0.8\textwidth]{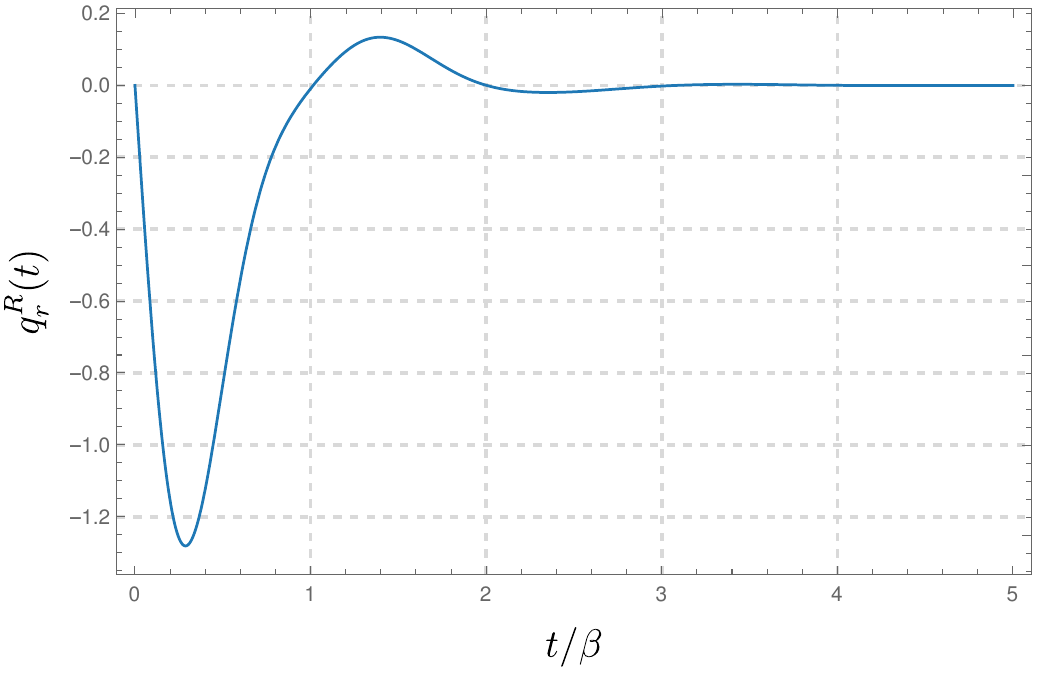}
        \caption{Large values of $M J$.}
    \end{subfigure}
    \caption{Representation of $q^R_r(t)$ as a function of time in the cases where $MJ$ are small and large.}
    \label{fig:qR_reps}
\end{figure}

We now return to the quantum spherical $p$-spin model. Applying the analytic continuation described above to the Euclidean saddle-point equation for $\hat{q}_r(k)$, we define the real-time retarded correlator $q_r^R(\omega)$, and its inverse Fourier transform, $q^R_r(t)$. In terms of the latter, the equations of motion read
\begin{equation}
	\frac{1}{q_r^R(\omega)}-\frac{1}{q_r^R(0)}=M \omega^2-J^2(\Lambda_r^R(\omega)-\Lambda_r^R(0)),
\end{equation}
where
\begin{equation}
	\Lambda_r^R(\omega)-\Lambda_r^R(0)=-\frac{ip}{2}\int_0^\infty dt(e^{-i\omega t}-1)\left[(q_r^>(t)+u)^{p-1}-(q_r^>(-t)+u)^{p-1}\right]\, .
\end{equation}

These equations can be solved numerically with a procedure analogous to that employed in the Euclidean case in Section \ref{sec:sphericalpspin}. In Fig. \ref{fig:qR_reps} we display the behavior of  $q^R(t)$ for small and large values of $MJ$. As this parameter grows and the system approaches a spin-glass regime, the dynamics undergoes a qualitative change: damped oscillations give way to a superexponential relaxation.

\section{Spin glasses, cosmology and holography}\label{sec:relationcosmo}

Up to this point, our focus has been on the construction and characterization of spin glass models as well-defined statistical and quantum systems. We now turn to a broader perspective and explain how similar ideas arise in cosmology and holography, even though spin glasses do not appear explicitly in those settings. In particular, we will focus on \textit{cosmic clustering} and \textit{AdS fragmentation}.

\paragraph{\bf Cosmology and cluster decomposition principle}

A basic expectation in quantum field theory is the cluster decomposition principle: observables measured at widely separated points in space should become uncorrelated \cite{Benincasa:2022gtd-ch05}. However, this principle does not generally hold in expanding universes, since two-point functions for massless fields grow logarithmically with spatial separation. Likewise, the quantum fluctuations of a scalar state might become large compared to the horizon scale and freeze in a branch diffusion process.

As a result, a given mode does not acquire a concrete value: as it evolves from the Bunch-Davies vacuum, it can freeze in many different late-time configurations. In this situation, it is natural to introduce a notion of distance between different late-time field configurations. Given two different field configurations $\phi_1(\eta_0,\vec{x})$ and $\phi_2(\eta_0, \vec{x})$, their mean square displacement is defined as \cite{Anninos:2011kh}
\begin{equation}
    \delta^{(12)} = d^{(12)}-\braket{d^{(12)}}\, ,
\end{equation}
with
\begin{equation}
    d^{(12)}=\frac{f}{L^d}\int d^d x \left[\hat{\phi}_1(\eta_0,\vec{x})-\hat{\phi}_2(\eta_0,\vec{x})\right]^2\,.
\end{equation}
Here $f$ is a function depending on the characteristics of the problem and $\hat{\phi}_i$ denotes the field configuration with a UV cut-off after the zero modes have been eliminated. Thanks to this definition of distance, it is possible to define a probability distribution of distances between two or more configurations $\mathcal{P}(\Delta_1,\Delta_2,...)$. In the previous function, $\Delta_i$ represent the distances between the different configurations involved.

Although a complete derivation of this distribution is out of the scope of these lectures, we want to mention that, remarkably, these distributions exhibit properties familiar from spin glass theory. In particular, the conditional probability $\mathcal{P}(\Delta_3|\Delta_1,\Delta_2)$ is peaked on configurations satisfying
\begin{equation}
    \Delta_3=\max(\Delta_1,\Delta_2)\, ,
\end{equation}
a geometric property known as ultrametricity and described originally in the context of spin glasses \cite{Parisi:1983dx,Rammal:1986zz}. This structure implies that the space of configurations can be organized as a tree, where each leaf (or end) corresponds to a given configuration and distances between them are measured as the depth of their common branch.

This is directly analogous to the structure of states in mean-field spin glasses, where ultrametricity emerges naturally from replica symmetry breaking. In this analogy, the distance distribution between cosmological configurations plays a role similar to that of the overlap matrix $Q_{ab}$ introduced in the replica formalism.

%\rafa{Not sure about how to face this first part} In models of universes characterized by an exponential expansion, the wave function describing it continuously dividing into different branches. Inside each branch, correlators at large separation factorize, but this phenomenon does not occur for the full quantum state. This process might be understood as a branched diffusion, where the quantum fluctuations of the light fields grow to super-horizon scales and become classical, in such a way that they are no longer in causal contact among each other.

%To see its relation with spin glasses, it is necessary to introduce a notion of distance between two different field configurations $\hat{\phi}_1(\vec{x}),\ \hat{\phi}_2(\vec{x})$ on a constant time slice
%\begin{equation}
%	d_{12}=\int d^dx (\hat{phi}_1(\vec{x})-\hat{phi}_2(\vec{x}))^2,
%\end{equation}
%which induces a distance distribution between fields. Its derivation is out of the scope of this lecture, but we highlight the fact that this function displays \textit{ultrametricity}\rafa{I have to complete it}

\paragraph{\bf AdS fragmentation and disordered quantum mechanics}

Spin glass ideas had also appeared in holography, most notably in the context of \textit{AdS fragmentation instabilities} \cite{Maldacena:1998uz,Anous:2021eqj}. These arise when a single $AdS$ throat splits into multiple pieces through quantum tunneling, with each resulting throat developing its own near-horizon region \cite{Majumdar,Papaetrou:1947ib}. Considerable effort has been devoted to unravel the microscopic mechanism underlying this phenomenon.

To understand why spin glass models were proposed as candidates to shed some light on black hole fragmentation, it is useful to look at black hole realizations within string theory. This phenomenon appears for instance in multi-centered solutions to $\mathcal{N}=2$ supergravity in 4 dimensions, where each of the fragmented black holes exhibits an $AdS_2$ near-horizon region \cite{Denef:2000nb,LopesCardoso:2000qm,Bates:2003vx}. In suitable limits, these systems are dual to brane models known as quiver quantum mechanics (QQM) \cite{Douglas:1996sw,Denef:2002ru,Denef:2007vg}. Although these models were later shown not to exhibit a genuine spin glass phase \cite{Biggs:2023mfn}, it is still worthwhile to understand the original motivations that suggested spin glass behavior could emerge in these systems.

The salient features of these systems are, first, that they exhibit $\mathcal{N}=4$ supersymmetry, which coincides with the amount preserved by the corresponding black hole geometries; and second, they contain chiral multiplet degrees of freedom $(\phi_i^\alpha,\ \psi_i^\alpha,\ F_i^\alpha)$ as well as vector multiplet degrees of freedom $(\{A,\vec{\mathbf{X}},\lambda, D\}$. A crucial point is that the interactions among chiral fields are governed by a superpotential $W$, which is a polynomial function of the chiral bosons. The lowest order interaction term involving the $\phi$ fields is cubic and has the form

\begin{equation}
    W(\phi)\supset\Omega_{ijk}\phi^1_i\phi_j^2\phi_k^3.
\end{equation}
The coefficients $\Omega_{ijk}$ are related to the geometry of the manifold where the compactification is performed, and can be understood as a source of disorder. 

The potential that describes the behavior of the chiral interactions is given by
\begin{equation}
    V=\displaystyle \sum_{i,a} \left|\frac{\partial W}{\partial \phi^a_i}\right|^2+\frac{1}{g_{YM}^2} \sum_a \left(\theta_a -\sum_{i} |\phi_i^a|^2\right)^2 +\left(\frac{\partial^2 W}{\partial \phi_i^a\partial\phi_j^b}\psi_i^a\epsilon \psi_j^b+ \textup{h.c.}\right),
\end{equation}
where $\theta_a$ are the so-called Fayet-Illiopoulos parameters, which on the ground state satisfy 
\begin{equation}\label{eq:Fayet condition}
    \displaystyle\sum_i |\phi_i^a|^2=\theta_a,\quad \forall a.
\end{equation}

If we further take into account that fermions do not interact directly with each other but just through bosonic many-body interactions, the resemblance between this model and the $p$-spin glass model that we have described in detail along this lecture becomes manifest. In the latter, we have $N$ bosonic degrees of freedom interacting through
\begin{equation}
    V= J_{i_1 \cdots i_p} \sigma_{i_1}\cdots\sigma_{i_p}.
\end{equation}
Similarly, the condition \eqref{eq:Fayet condition} closely parallels the spherical constraint in \eqref{eq: spherical constraint}.\\

From this perspective, the glassy structure uncovered in the spherical $p$-spin model seemed to provide a useful toy framework for studying semiclassical black holes and their fragmentation. While the correspondence was not exact, the shared features between both types of systems suggested that techniques from spin glass theory could provide valuable progress in the understanding of the microscopic origin of the black hole fragmentation (see for instance, \cite{Anninos:2011vn,Anninos:2012gk,Anninos:2013nra,Anninos:2013mfa,Kent-Dobias:2020egr}).

\section{Summary}\label{sec:conclusions}

In these notes we have presented a detailed study of the spherical $p$-spin model. Starting from a simple mean-field model, we introduced the basic ingredients that distinguish glassy systems from more conventional phases of matter, such as frustration and disorder. The spherical constraint allowed for a controlled large-$N$ limit analysis, making it possible to derive exact saddle-point equations for the thermodynamics of the model.

Using the replica method, we characterized both the replica-symmetric phase and the glassy phase described by one-step replica symmetry breaking. The latter is characterized by a nontrivial overlap structure, encoded in the Edwards-Anderson order parameter and the breaking parameter $m$.

We then extended the analysis to the quantum spherical $p$-spin model, where quantum fluctuations are controlled by a kinetic term. In this model, the saddle-point equations of motion become dynamical and are naturally formulated in terms of correlation functions. By studying their Euclidean solutions and performing an analytic continuation to real time, we obtained a description of the system's quantum dynamics across the paramagnetic and spin glass regimes. The resulting real-time correlators exhibit a qualitative change in behavior as the system enters the glassy phase, reflecting the slowing down and eventual freezing of dynamics characteristic of glasses.

Finally, we discussed how the ideas and techniques developed in these notes extend beyond their original condensed matter context. Concepts such as replica symmetry breaking, ultrametricity, and disordered many-body interactions have found applications in cosmology and holography, where they provide useful tools for understanding phenomena such as cosmological clustering and AdS fragmentation. While spin glass models are not literal descriptions of this system, they offer tractable models that capture such essential features.

These notes are intended as a self-contained introduction to these ideas, providing both technical tools and conceptual motivation. We hope they serve as a useful starting point for readers interested in the rich connections between spin glasses, quantum dynamics, and modern problems in theoretical physics.

\appendix

\begin{acknowledgements}
    -- We would like to thank Paolo Benincasa and Guilherme Pimentel for organizing this excellent school, as well as for their hospitality throughout the event. We are also deeply grateful to Ana, Borja and Paula for their tireless efforts in ensuring everything ran smoothly. We also want to thank Dionysios Anninos, Tarek Anous, Thomas Colas, and Andrew McLeod for their engaging and insightful lectures. Finally, we want to thank all the participants for fostering a friendly atmosphere and for the many fruitful discussions.
    
    The \href{https://indico.sns.it/event/91/}{{\it La Ricotta} Summer School  {\it The Disordered Universe 2025}} received financial support from the ERC grant NOTIMEFORCOSMO (agreement n${}^{\circ}$ 1126304)), by Scuola Normale and by the Instituto Galego de F{\'i}sica das Altas Enerx{\'i}as (IGFAE-USC/Xunta de Galicia) through the CIGUS network and EU cofinancing program.
\end{acknowledgements}

\section{Solutions to the Exercises}\label{app:solutions}
In this Appendix we provide the solutions to the proposed exercises throughout the main text.

\subsection{Exercises from Section \ref{sec:meanfield}}

\printexercise{ex:integrateoutz}

\textbf{Solution: }In order to integrate out the variable $z$ in \eqref{eq:ZN-zm01}, we will make use of the steepest descent method. This method establishes that the integral of this exponential is approximately equal to its value at the saddle. We will write the partition function as
$$
Z_N= \frac{1}{2\pi}\sqrt{\frac{N\beta J}{2\pi}}\int_{-\infty}^{\infty} \text{d} z \int_{-\infty}^{\infty} \text{d} m \exp \left(\frac{N}{2} \Phi(m, z)\right)
$$
where 

$$\Phi(m,z)=-\beta J m^2 + \frac{1}{2}\frac{(h+\beta Jm)^2}{a+iz +\frac{\beta J}{N}}\\
    +2(a+iz)-\log \frac{a+iz+\frac{\beta J}{N}}{\pi}.$$
integrate out $z$ at large $N$ (steepest descent), and extract the effective free-energy functional. According to \eqref{eq:free_energy_part_function}, once the integration over $z$ has been performed, we can identify the free energy as the argument of the exponential and we can compute the critical temperature $T_c$.

The saddle point is obtained from the condition 
$$
\partial_x\Phi=-\frac{(h+\beta J m)^2}{2 x^2}+2-\frac{1}{x}=0\, .
$$
Multiplying and dividing by $2x^2$ one obtains
$$
4 x^2-2 x-(h+\beta J m)^2=0 \,.
$$
The resolution of the previous equation provides a pair of saddles
$$x_\star(m)=\frac{1\pm\sqrt{1+4(h+\beta Jm)^2}}{4}.$$
However, the dominant saddle will be the one that maximizes the argument of the exponential so, from these two, we will have to consider the positive one.

The (relevant) solution to this equation reads
$$
x_\star(m)=\frac{1+\sqrt{1+4(h+\beta Jm)^2}}{4}.
$$

In summary, after solving the $z$-saddle (equivalently $x=a+iz$) at fixed $m$, i.e. $x=x_*(m)$, and
substituting it back into the integrand, the $z$-integral gives at leading order in $N$ 
\[
Z_N(\beta,h)\;\approx\;\frac{1}{2\pi}\sqrt{\frac{N\beta J}{2\pi}} \int_{-\infty}^{\infty} dm\;
\exp\!\left[\frac{N}{2}\,\Phi\big(m,x_*(m)\big)\right].
\]
Since $\beta=1/T$, we can rewrite the exponent in the standard form $-\beta N f$ by defining
$f(m,T,h)$ through
\[
\exp\!\left[\frac{N}{2}\,\Phi\big(m,x_*(m)\big)\right]
\;\equiv\;\exp\!\big[-\beta N f(m,T,h)\big],
\]
that is,
\[
f(m,T,h)\;\equiv\;-\frac{T}{2}\,\Phi\big(m,x_*(m)\big)\,.
\]

More explicitly, after plugging this solution into the definition of the partition function, it is found
\begin{equation}
\begin{aligned} \label{eq:ZNex1a}
    Z_N[\beta,h]=&\frac{1}{2\pi}\sqrt{\frac{N \beta J}{2\pi}} \int_{-\infty}^{\infty}\text{d}m\exp
\Bigg(\frac{N}{2}\Bigg(-\beta J m^2 + \frac{2(h+\beta Jm)^2}{1+\sqrt{1+4(h+\beta J m)^2}}\\
   & +\frac{1+\sqrt{1+4(h+\beta J m)^2}}{2}-\log \frac{1+\sqrt{1+4(h+\beta J m)^2}}{4\pi}\Bigg)\Bigg)\, ,
    \end{aligned}
\end{equation}
so that $f(m,T,h)$ reads
\begin{equation}
\begin{split}
    f(m,\beta,h)=&\frac{1}{2}\Bigg( J m^2 - \frac{2(h+\beta Jm)^2}{\beta(1+\sqrt{1+4(h+\beta J m)^2})}\\
   & -\frac{1+\sqrt{1+4(h+\beta J m)^2}}{2\beta}+\frac{1}{\beta}\log \frac{1+\sqrt{1+4(h+\beta J m)^2}}{4\pi}\Bigg)
\end{split}
\end{equation}
 Furthermore, we realize that setting $h= 0$, we obtain eq. \eqref{eq:freee}. Let us study in more detail the case in which $h=0$ and $m\rightarrow0$. Performing a series expansion around $m=0$ in the previous equation, we find ($\beta=1/T$)
\begin{equation}
    f(m,T,0)=-\frac{T}{2}\left(1+\log(2\pi)\right)+\frac{J}{2}\Big(1-\frac{J}{T}\Big)m^2+\frac{J^4}{4T^3}m^4+O(m^6).
\end{equation}

To obtain the critical temperature, we set $h=0$, and expand around $m=0$. Using $x_\star(m)$ (as computed above) one finds 
\begin{equation}
f(m,T,0)=f(0,T,0)+\frac{J}{2}\Big(1-\frac{J}{T}\Big)m^2+\frac{J^4}{4T^3}m^4+O(m^6).
\end{equation}
Hence the curvature at the origin changes sign at
\begin{equation}
T_c=J.
\end{equation}
For $T>T_c$ the minimum is at $m=0$ (single-well), while for $T<T_c$ the functional develops
two symmetric minima (double-well), corresponding to spontaneous magnetization.

\printexercise{ex:susceptibility}

\textbf{Solution: }From Exercise \ref{ex:integrateoutz} we have, at large $N$,
\[
Z_N(\beta,h)\simeq \int_{-\infty}^{\infty} \text{d}m\; \exp\!\big[-\beta N f(m,T,h)\big],
\]
so the magnetization is the saddle $m^*(T,h)$ determined by
\[
\left.\frac{\partial f}{\partial m}\right|_{m=m^*}=0.
\]
Using also the $z$-saddle equation from 1.a to eliminate $z$, this condition gives the
equation of state (same conventions as the notes, where $h$ enters as $+h\sum_i\sigma_i$):
\begin{equation}\label{eq:eos_1b}
h(m)=\frac{m}{1-m^2}-\beta J\,m .
\end{equation}

The susceptibility is
\[
\chi=\left.\frac{\partial m^*}{\partial h}\right|_{h=0}
=\left[\left.\frac{dh}{dm}\right|_{m=m_0}\right]^{-1},
\qquad
\frac{dh}{dm}=\frac{1+m^2}{(1-m^2)^2}-\beta J,
\]
where $m_0=\lim_{h\to0^+}m^*(T,h)$.

For $T>T_c$ the $h\to0$ solution is $m_0=0$, hence
\[
\chi=\frac{1}{1-\beta J}=\frac{T}{T-J}.
\]

For $T<T_c$ the $h\to0^+$ solution satisfies $h(m_0)=0$ with $m_0\neq 0$, i.e.
\[
\frac{1}{1-m_0^2}=\beta J \quad\Rightarrow\quad m_0^2=1-\frac{T}{J},\qquad (T_c=J),
\]
and therefore
\[
\left.\frac{dh}{dm}\right|_{m_0}
=\frac{1+m_0^2}{(1-m_0^2)^2}-\beta J
=2\beta J(\beta J-1),
\qquad
\chi=\frac{1}{2\beta J(\beta J-1)}
=\frac{T^2}{2J(J-T)}.
\]

In summary,
\[
\chi=
\begin{cases}
\displaystyle \frac{1}{2J^2}\,\frac{T^2}{1-\frac{T}{T_c}}, & T<T_c,\\[2mm]
\displaystyle \frac{1}{J}\,\frac{T}{\frac{T}{T_c}-1}, & T>T_c,
\end{cases}
\qquad T_c=J.
\]

\printexercise{ex:FSandE}

\textbf{Solution: }From Exercise \ref{ex:integrateoutz},
\[
Z_N(\beta,0)\simeq \frac{1}{2\pi}\sqrt{\frac{N\beta J}{2\pi}} \int \text{d}m\;\exp\!\big[-\beta N f(m,T,0)\big],
\]
so for $N\gg 1$ one can ignore the prefactor, since it is independent of $m$ and contributes only at $O (\log N/N)$ to the free energy $F/N$. Furthermore the $m$-integral is dominated by the saddle $m^*(T)$, and
%so for $N\gg 1$ the integral is dominated by the saddle $m^*(T)$ and
\[
\frac{F}{N}=-\frac{T}{N}\log Z_N \;\approx\; f\big(m^*(T),T,0\big)\,.
\]
Using the $h=0$ saddle solution from Exercise \ref{ex:susceptibility}, one has $T_c=J$ and
\[
m^*(T)=0 \quad (T>T_c), \qquad (m^*(T))^2=1-\frac{T}{T_c}=1-\frac{T}{J}\quad (T<T_c).
\]
% Evaluating the free-energy density at the saddle gives
% \[
% \frac{F}{N}=
% \begin{cases}
% 0, & T>J,\\[6pt]
% -\dfrac{J}{2}+\dfrac{T}{2}-\dfrac{T}{2}\log\!\Big(\dfrac{T}{J}\Big), & T<J.
% \end{cases}
% \]
% Equivalently, for $T<J$,
% \[
% \frac{F}{N}=-\frac{J}{2}-\frac{T}{2}\log\!\Big(\frac{T}{Je}\Big)
% = a+b\,T\log\frac{T}{c},
% \]
% so that
% \[
% T_c=J,\qquad a=-\frac{J}{2},\qquad b=-\frac12,\qquad c=Je,\qquad d=0.
% \]
%
% Now use the canonical identities (as in the hints):
% \[
% \frac{\langle E\rangle}{N}
% =\frac{F}{N}-T\frac{\partial}{\partial T}\!\left(\frac{F}{N}\right),
% \qquad
% \frac{S}{N}=-\frac{\partial}{\partial T}\!\left(\frac{F}{N}\right),
% \qquad
% \frac{C_T}{N}=\frac{\partial}{\partial T}\!\left(\frac{\langle E\rangle}{N}\right).
% \]
% From the piecewise $F/N$ above one obtains
% \[
% \frac{\langle E\rangle}{N}=
% \begin{cases}
% 0, & T>J,\\[6pt]
% -\dfrac{J}{2}+\dfrac{T}{2}, & T<J,
% \end{cases}
% \qquad
% \frac{C_T}{N}=
% \begin{cases}
% 0, & T>J,\\[4pt]
% \dfrac12, & T<J,
% \end{cases}
% \]
% so the constant in the statement is $\#=N/2$.
%
% \medskip
%
Evaluating the free-energy density at the saddle gives 
\[
\frac{F}{N}=
\begin{cases}
-\dfrac{T}{2}\big(1+\log(2\pi)\big), & T>J,\\[6pt]
-\dfrac{J}{2}+\dfrac{T}{2}\log\!\Big(\dfrac{J}{2\pi T}\Big), & T<J.
\end{cases}
\]
Equivalently, for $T<J$, 
\[
\frac{F}{N}
=-\frac{J}{2}-\frac{T}{2}\log\!\Big(\frac{T}{J/(2\pi)}\Big)
= a+b\,T\log\frac{T}{c},
\]
and for $T>J$ we have $\frac{F}{N}=d\,T$, so that
\[
T_c=J,\qquad a=-\frac{J}{2},\qquad b=-\frac12,\qquad c=\frac{J}{2\pi},
\qquad d=-\frac12\big(1+\log(2\pi)\big).
\]

\medskip
Now use the canonical identities (as in the hints):
\[
\frac{\langle E\rangle}{N}
=\frac{F}{N}-T\frac{\partial}{\partial T}\!\left(\frac{F}{N}\right),
\qquad
\frac{S}{N}=-\frac{\partial}{\partial T}\!\left(\frac{F}{N}\right),
\qquad
\frac{C_T}{N}=\frac{\partial}{\partial T}\!\left(\frac{\langle E\rangle}{N}\right).
\]
From the piecewise $F/N$ above one obtains
\[
\frac{\langle E\rangle}{N}=
\begin{cases}
0, & T>J,\\[6pt]
-\dfrac{J}{2}+\dfrac{T}{2}, & T<J,
\end{cases}
\qquad
\frac{C_T}{N}=
\begin{cases}
0, & T>J,\\[4pt]
\dfrac12, & T<J,
\end{cases}
\]
so the constant in the statement is $\#=N/2$.
With this convention,
\[
\frac{S}{N}=
\begin{cases}
\dfrac12\big(1+\log(2\pi)\big), & T>J,\\[6pt]
\dfrac12\log\!\Big(\dfrac{2\pi \,T}{J}\Big)+\dfrac12, & T<J.
\end{cases}
\]

%With this normalization, $S/N=0$ for $T>J$ and $S/N=\frac12\log(T/J)$ for $T<J$.

\subsection{Exercises from Section \ref{sec:sphericalpspin}}

\printexercise{ex:approximation}

\textbf{Solution:} Consider some function $f$ of the indices such that $f(i_1,...,i_p)$ are of order $\mathcal{O}(1)$ at large $N$. The expression
\begin{equation}
    \label{sum of all}
    \sum_{i_1,...,i_{p}=1}^{N}f(i_1,...,i_p)
\end{equation}
sums over all $N^p$ terms, while the expression
\begin{equation}
    \label{distinct terms}
    p!\sum_{1\leq i_1<...<i_{p}\leq N}f(i_1,...,i_p)
\end{equation}
sums only over the terms with distinct indices. Counting only the ordered ones explains the $p!$ factor. The remaining terms are those where at least two indices coincide. For instance, there are $\mathcal{O}(N^{p-1})$ terms where exactly two indices coincide, $\mathcal{O}(N^{p-2})$ terms where exactly three coincide, and so on. All such terms can be neglected at large $N$ compared to the $\mathcal{O}(N^p)$ terms where no indices coincide.

\printexercise{ex:RSansatz}

\textbf{Solution:} Observe that the matrix in \eqref{replica-symmetric Q} can be written as $Q_{RS}=(1-u)I+uJ$ where $I$ is the $n \times n$ identity matrix and $J$ is a matrix with all coefficients equal to one. Then, any eigenvector of $J$ is also an eigenvector of $Q_{RS}$. For instance, the obvious $v_1=(1,\ldots,1)$ eigenvector carries the eigenvalue $(1-u)+nu$ for $Q_{RS}$. Moreover, given that $J$ has rank $1$, it annihilates all other vectors linearly independent with $v_1$. And so, these $n-1$ eigenvectors all carry an eigenvalue $1-u$.

\printexercise{ex:RSBsolution}
\textbf{Solution: }
\begin{enumerate}
 \item The matrix is $n\times n$ with only diagonal entries 1, so the total number of $1$'s is $n$. Each $m\times m$ block has $m$ diagonal entries equal to one, while the rest are $u$. Hence, within a block, there are $m^2-m$ entries equal to $u$, and there are $n/m$ such blocks. This makes the total number of $u$'s to be $n(m-1)$. Finally, all other entries are zero, so the number of $s$'s is $n^2-n-n(m-1)$, which simplifies to $n(n-m)$.
 \item A simple way to obtain the eigenvalues is to guess the eigenvectors. The simplest guess is the vector with all entries set to one. This guess works because each row of the matrix has a single entry equal to $1$, $(m-1)$ entries equal to $u$ and $n-m$ entries equal to $s$, giving the eigenvalue for this vector: $1+(m-1)u+(n-m)s$. This corresponds to the sum over an entire row(or equivalently, an entire column) of entries, and the fact that it is the same for every row(column) is enforced by the equivalence of the replicas. 

In the next guess, we can exploit the matrix's block structure. A vector with first $m$ entries equal to $1$, the next $m$ entries equal to $-1$ and the rest of the entries set to zero, gives eigenvalue $1+u(m-1)-ms$. This exploits the structure of adjacent blocks, and hence, one can find such an eigenvector for each pair of neighbouring blocks. Since there are $n/m$ blocks in total, there are $n/m-1$ total adjacent pairs, each providing an eigenvector with eigenvalue $1+u(m-1)-ms$. Hence, the multiplicity of the eigenvalue is $n/m-1$.

Finally, we can borrow an eigenvector from the solution of the replica symmetric ansatz, i.e. one with an adjacent pair of $1$ and $-1$ with other entries set to zero. There are $m-1$ such entries per $m\times m$ block, and there are $n/m$ such blocks, which makes the multiplicity equal to $n(1-1/m)$. The eigenvalue can be checked to be $1-u$ for these eigenvectors. Adding up all the multiplicities of the eigenvectors we have found till now gives $n$, and hence, we have classified all possible eigenvalues. 
\end{enumerate}

\subsection{Exercises from Section \ref{sec:dynamics}}

\printexercise{ex:eomqr}

\textbf{Solution: }
Varying the action \eqref{eq:Sfinal} one obtains
\begin{multline}
\label{eq:eomq}
\text{EOM}_q:\qquad	M\left(\frac{2\pi k}{\beta}\right)^2-2 i\frac{\hat{z}(0)}{\beta}=\frac{\hat{q}(k)+\beta\,u(m-2)\delta_{k,0}}{\hat{q}(k)^2-\beta\,u\,\delta_{k,0}\left[(m-1)\beta\,u-(m-2)\hat{q}(k)\right]}\\+p\frac{\beta^2J^2}{2}\sum_{k_2,\dots,k_p}\frac{\hat{q}(k_2)\dots\hat{q}(k_p)}{\beta^p}\delta(k+k_2+\dots+k_p)~.
\end{multline}
Using the definition of the self-energy $\Lambda(\tau,\tau')$ as in \eqref{eq:Lambdadef},
\begin{equation}
    \Lambda(\tau,\tau')\equiv \frac{p}{2}q(\tau,\tau')^{p-1},
\end{equation}
whose Fourier transform is
\begin{equation} 
\hat{\Lambda}(k) = \frac{p}{2\beta^{p-2}} \sum_{k_1,\dots,k_{p-2}} \hat{q}(k_1)\dots\hat{q}(k_{p-2})\hat{q}(k-k_1-\dots-k_{p-2})~,
\end{equation}
we can rewrite \eqref{eq:eomq} as
\begin{equation}
\label{eq:EOMsigma}
	\frac{\hat{q}(k)+\beta\,u(m-2)\delta_{k,0}}{\hat{q}(k)^2-\beta\,u\,\delta_{k,0}\left[(m-1)\beta\,u-(m-2)\hat{q}(k)\right]}= M \left(\frac{2\pi k}{\beta}\right)^2-2 i   \frac{\hat{z}(0)}{\beta} - J^2 \, \hat{\Lambda}(k) \,.
\end{equation}
We can use now the definitions given in \eqref{eq:subtractSGparams}, {\it i.e.},
\begin{equation}
\begin{split}
   q_r(\tau,\tau') &= q(\tau,\tau') - u \,,\qquad \Lambda_r(\tau,\tau') = \Lambda(\tau,\tau') - \frac{p}{2}\, u^{p-1} \,,\\
   \hat{q}_r(k) &= \hat{q}(k) - \beta u \delta_{k,0} \,,\qquad\;
   \hat{\Lambda}_r(k) = \hat{\Lambda}(k) - \frac{p}{2} \beta u^{p-1} \delta_{k,0}
   \end{split}
\end{equation}
with boundary conditions $q_r(\tau,\tau) = 1-u$ and $\Lambda_r(\tau,\tau) = \frac{p}{2} (1-u^{p-1})$.
The equation of motion \eqref{eq:EOMsigma} then becomes very simple for all values of $k$:
\begin{equation}
\label{eq:EOMsigma2}
\text{EOM}_q:\qquad	\frac{1}{\hat{q}_r(k)}= M \left(\frac{2\pi k}{\beta}\right)^2-2   \frac{i\hat{z}(0)}{\beta} - J^2 \, \hat{\Lambda}_r(k) \,,
\end{equation}
where we used \eqref{eq:eomuq} to cancel the terms proportional to $\delta_{k,0}$. Evaluating this for $k=0$, we can solve for the Lagrange multiplier $\hat{z}(0)$:
\begin{equation}
\label{eq:zSol}
 	\frac{-i\hat{z}(0)}{\beta}\equiv \frac{1}{2} \left[ \frac{1}{\hat{q}_r(0)} + J^2 \hat{\Lambda}_r(0)\right]=\frac{1}{2}\left[\frac{1}{\int_0^\beta\, d\tau\, q_r(\tau)}+J^2\int_0^\beta d\tau\, \Lambda_r(\tau)\right] ~ .
 \end{equation} 
Combined with \eqref{eq:EOMsigma2} we obtain the desired expression for the equation of motion for $\hat{q}_r(k)$:
\begin{equation}
\label{eq:EoMq}
\boxed{ \;\;\text{EOM}_q:\qquad 	\frac{1}{\hat{q}_r(k)}-\frac{1}{\hat{q}_r(0)}= M \left(\frac{2\pi k}{\beta}\right)^2- J^2 \, (\hat{\Lambda}_r(k)-\hat{\Lambda}_r(0)  ) \, .
 }\end{equation}

\bibliographystyle{utphys}
\bibliography{2025-DU-SSOE-biblio}

\end{document}